\documentclass[twocolumn]{aastex631}

\usepackage{xcolor}
\usepackage{amsmath}
\usepackage{hyperref}
\usepackage{xspace}
\usepackage{natbib}
\usepackage{graphicx}
\graphicspath{{./figures/}{../figures/}{./}}

\newcommand{\Lya}{Ly$\alpha$}
\newcommand{\AAGN}[1]{$A_\textnormal{AGN#1}$}
\newcommand{\ASN}[1]{$A_\textnormal{SN#1}$}

\shorttitle{Feedback effects on low-$z$ \Lya~forest with CAMELS}
\shortauthors{Tillman et al.}

\begin{document}

\title{An Exploration of AGN and Stellar Feedback Effects in the Intergalactic Medium via the  Low Redshift Lyman-$\alpha$ Forest}

\author[0000-0002-1185-4111]{Megan Taylor Tillman}
\affiliation{Department of Physics and Astronomy, Rutgers University,  136 Frelinghuysen Rd, Piscataway, NJ 08854, USA}

\author[0000-0001-5817-5944]{Blakesley Burkhart}
\affiliation{Department of Physics and Astronomy, Rutgers University,  136 Frelinghuysen Rd, Piscataway, NJ 08854, USA}
\affiliation{Center for Computational Astrophysics, Flatiron Institute, 162 Fifth Avenue, New York, NY 10010, USA}

\author[0000-0002-8710-9206]{Stephanie Tonnesen}
\affiliation{Center for Computational Astrophysics, Flatiron Institute, 162 Fifth Avenue, New York, NY 10010, USA}

\author[0000-0001-5803-5490]{Simeon Bird}
\affiliation{University of California, Riverside, 92507 CA, U.S.A.}

\author[0000-0003-2630-9228]{Greg L. Bryan}
\affiliation{Center for Computational Astrophysics, Flatiron Institute, 162 Fifth Avenue, New York, NY 10010, USA}
\affiliation{Department of Astronomy, Columbia University, 550 W 120th Street, New York, NY 10027, USA}

\author[0000-0001-5769-4945]{Daniel Angl\'es-Alc\'azar} 
\affiliation{Department of Physics, University of Connecticut, 196 Auditorium Road, U-3046, Storrs, CT 06269-3046, USA}
\affiliation{Center for Computational Astrophysics, Flatiron Institute, 162 Fifth Avenue, New York, NY 10010, USA}

\author[0000-0002-1050-7572]{Sultan Hassan}
\affiliation{Center for Cosmology and Particle Physics, Department of Physics, New York University, 726 Broadway, New York, NY 10003, USA}
\affiliation{Center for Computational Astrophysics, Flatiron Institute, 162 Fifth Avenue, New York, NY 10010, USA}\affiliation{Department of Physics \& Astronomy, University of the Western Cape, Cape Town 7535,
South Africa}

\author[0000-0002-6748-6821]{Rachel S. Somerville}
\affiliation{Center for Computational Astrophysics, Flatiron Institute, 162 Fifth Avenue, New York, NY 10010, USA}

\author[0000-0003-2842-9434]{Romeel Dav\'e}
\affiliation{Institute for Astronomy, University of Edinburgh, Royal Observatory, Edinburgh, EH9 3HJ, UK}
\affiliation{University of the Western Cape, Bellville, Cape Town 7535, South Africa}

\author[0000-0003-3816-7028]{Federico Marinacci}
\affiliation{Department of Physics and Astronomy ''Augusto Righi", University of Bologna, Via P. Gobetti 93/2, I-40129 Bologna, Italy}

\author{Lars Hernquist}
\affiliation{Center for Astrophysics | Harvard \& Smithsonian, 60 Garden St., Cambridge, MA 02138, USA}

\author[0000-0001-8593-7692]{Mark Vogelsberger}
\affiliation{Department of Physics, Kavli Institute for Astrophysics and Space Research, Massachusetts Institute of Technology, Cambridge, MA 02139, USA}

\begin{abstract}
   We explore the role of galactic feedback on the low redshift Lyman-$\alpha$ (\Lya)~forest ($z \lesssim 2$)  statistics and its potential to alter the thermal state of the intergalactic medium. 
   Using the Cosmology and Astrophysics with Machine Learning Simulations (CAMELS) suite,  we explore variations of the AGN and stellar feedback models in the IllustrisTNG and Simba sub-grid models. 
   We find that both AGN and stellar feedback in Simba play a role in setting the \Lya~forest column density distribution function (CDD) and the Doppler width ($b$-value) distribution.
   The Simba AGN jet feedback mode is able to efficiently transport energy out to the diffuse IGM causing changes in the shape and normalization of the CDD and a broadening of the $b$-value distribution. 
   We find that stellar feedback plays a prominent role in regulating supermassive black hole growth and feedback, highlighting the importance of constraining stellar and AGN feedback simultaneously.
   In IllustrisTNG, the AGN feedback variations explored in CAMELS  do not affect the \Lya~forest, but varying the stellar feedback model does produce subtle changes.
   Our results imply that the low-$z$ \Lya~forest can be sensitive to changes in the ultraviolet background (UVB), stellar and black hole feedback, and that AGN jet feedback in particular can have a strong effect on the thermal state of the IGM.
\end{abstract}

\keywords{Cosmology; Extragalactic astronomy; Intergalactic gas; Intergalactic medium;
Lyman alpha forest; Active galactic nuclei; Supermassive black holes; Stellar feedback}

\section{Introduction}\label{s:Introduction}

    Over the past 50 years, the \Lya~forest has been used at high redshift ($z>2$) to constrain the amplitude of cosmological fluctuations, the thermal properties of the intergalactic medium (IGM), and the process of reionization \citep[see][for a review]{McQuinn:2016}. 
    The majority of matter in the universe resides within the IGM, the space between galaxies, making the absorption lines that make up the forest a powerful tool for probing otherwise unobservable gas.
    An understanding of radiative processes reveals from the \Lya~forest spectra the column density of the absorbing structure as well as its thermal properties and kinetics properties. 
    The observed forest has been used to study small-scale cosmic structure \citep{Croft:1998,McDonald:2005}, the temperature of dark matter \citep{Viel:2009},  gas temperature \citep{Becker:2011, Boera:2014}, the evolution of the metagalactic ionizing background \citep{Fan:2006,Faucher-Giguere:2009,Haardt:2012}, and the mapping of neutral hydrogen \citep{Lee:2016, Ozbek:2016}.
    Additionally, simulations of the forest have enabled constraints on baryons, dark matter interactions, and  photoionization heating and adiabatic cooling in the IGM \citep{Cen:1994,Zhang:1995,Miralda-Escude:1996,Hernquist:1996,Rauch:1997}. 
    In more recent years, efforts with the high-$z$ \Lya~forest data have led to further constraints on ionizing background models and the termperature of the IGM \citep{Gaikwad:2017,Onorbe:2017,Chardin:2017,DAloisio:2017,walther:2018,Puchwein:2019,FG:2020}, properties of dark matter \citep{Irsic:2017,Rogers:2021}, the time frame and processes of the epoch of HI and HeII reionization  \citep{Zhu:2021,Gaikwad:2021,Villasenor:2022,Wang:2022,Yang:2023}, and the 3D mapping of cosmic structures \citep{Lee:2018,Horowitz:2022,Quezlou:2022,Newman:2022}.
    
    When utilizing the high redshift IGM to constrain cosmology, the role of astrophysical processes can impact predictions on the linear matter power spectrum \citep{Pritchard:2007,Wyithe:2011}, and the use of eBOSS data is sufficiently constraining to break most of these degeneracies \citep{Bird:2011}. 
    Additionally, localized fluctuations of the ionizing background or inhomogeneous reionization can impact the observed \Lya~forest flux power spectrum leading to bias in predictions of the matter power spectrum \citep{McQuinn:2016}. 
    These challenges require studies to be mindful of the impact that astrophysical processes, including feedback, can have on predictions that are based on the \Lya~forest.
    However, radiation backgrounds are often approximated as uniform for large cosmological simulations so the effects of photoionization has largely been included via models for the ultraviolet background (UVB).
    These uniform ionizing background models include ionizing photons from galactic feedback by assuming  some fraction of the photons come from feedback by AGN and supernovae.
    
    Despite the robustness of high-$z$ IGM studies, the low redshift ($z<2$) IGM has received considerably less attention, with one early exception being the work of \cite{dave:1999}. At low-$z$ the \Lya~forest is observed in the far-ultraviolet (FUV) band, requiring the use of space-based telescopes.
    Only a few instruments such as the Hubble Space Telescope's (HST) Cosmic Origins Spectrograph (COS) and the Far Ultraviolet Spectroscopic Explorer (FUSE) satellite can observe in the UV band and collect \Lya~data \citep{Danforth:2005}.
    In particular, HST COS can observe \Lya~from $z\approx0-0.5$ and the \citeauthor{Danforth:2016}~(\citeyear{Danforth:2016}; henceforth D16) catalog has built on previous surveys to produce the largest collection of low-$z$ \Lya~forest absorber data to date. This allows for the exploration of the low-$z$ \Lya~forest in a way previously unattainable. 
    
    Since the D16 study, many theoretical efforts have explored the low-$z$ \Lya~forest due to its relatively untapped potential.
    In particular, the role of feedback in the low redshift \Lya~forest remains relatively unclear. \citet{Kollmeier:2014} reported a disconnect between the observed and simulated \Lya~forest column density distribution function (CDD), which required either an HI photoionizing rate a factor of 5 times higher than UVB models predicted at the time or the existence of nontraditional heating sources. While the factor of 5 may be partly attributed to poor resolution, it was later shown that a large increase in the UVB magnitude from \citet{Haardt:2012} (the UVB model studied in \citealt{Kollmeier:2014}) is possible \citep{Khaire:2015,Puchwein:2019,Bolton:2021}.
    As for additional heating sources, several studies explore the potential impact of AGN feedback in heating the IGM, with some finding that different AGN feedback models can have a dramatic impact on the low-$z$ \Lya~forest statistics \citep{Gurvich2017,Nasir:2017,Christiansen:2020,Burkhart_2022,Tillman:2023,Khaire:2023}. 
    Fortunately, modern day simulations have put a greater emphasis on incorporating feedback mechanisms through sophisticated modeling based on observations \citep[e.g.,][and \citealt{Vogelsberger:2020} for a review]{Illustris,vogelsberger:2014b,hopkins+2014,hopkins+2018,EAGLE, IllustrisTNG, Kaviraj:2017,SIMBA,Bird:2022} making the study of feedback effects on the low-$z$ forest possible through simulated data.

    While the effects of some AGN feedback models on the statistics of the forest are degenerate with a re-scaling of the assumed UVB \citep[e.g.\ see][]{Burkhart_2022, Khaire:2023, Mallik:2023}, other AGN models had different, idiosyncratic, effects, highlighting the importance of including feedback sub-grid models within cosmological simulations \citep{Tillman:2023}.
    The unique effects seen on the \Lya~forest column density distribution function (CDD) due to AGN jet feedback \citep{Tillman:2023} are primarily attributed to the far-reaching heating effects that the jet feedback has on the large-scale environment \citep[see][for discussion of the dependence of the CDD on the large-scale environment]{Tonnesen:2017}.
    These results emphasize the need for an improved knowledge of the physical conditions of the IGM and how they vary on large scales.
    This not only includes a better understanding of UV ionizing photon sources, but also any astrophysical effects such as feedback that might affect the HI within the IGM via heating or gas transport. 

    The \Lya~forest Doppler width, or $b$-value, distribution can provide information about the thermal and turbulent state of absorbers.
    However, matching the simulated $b$-value distribution with the observed distribution has proved more difficult than matching the CDD. 
    The simulations consistently under-predict the number of high $b$ absorbers as compared to observations. 
    Several studies have posed the idea of introducing non-traditional heating sources, such as galactic feedback, or extra line of sight turbulent velocity components in simulations to alleviate the tension \citep{Viel:2017, Gaikwad:2017viper, Bolton:2021, Burkhart_2022}.
    \citet{Bolton:2021} and \citet{Viel:2017} both discuss the difficulties of using AGN feedback as a solution to correcting the $b$-value distribution. 
    The AGN feedback models explored in these studies (which were heavily based on the Illustris and IllustrisTNG models) failed to reproduce the observed $b$-value distribution, however, whether this statistic might be used to constrain other feedback models is unclear. 
    It is likely that the combination of unresolved turbulence and heating from galactic feedback that does not over-ionize HI in the IGM is required to reproduce the observed $b$-value distribution. 

    To further understand the role of AGN feedback in the low-$z$ \Lya~CDD and $b$-value distributions, a systematic exploration of multiple different AGN feedback models is required. 
    The CAMELS simulations provide an excellent opportunity for this type of analysis \citep{camels}. 
    By varying multiple parameters for both AGN and stellar feedback we can not only explore the effects of AGN feedback models in different simulations, but we can also explore the variation of feedback parameters within each model.

    In this study, we explore variations on the Simba and IllutrisTNG stellar and AGN feedback models using the CAMELS project simulations. 
    We analyze both the \Lya~forest CDD and $b$-value distributions within the context of these simulations.
    In Section \ref{s:Methods} we describe the CAMELS project and provide a brief description of the relevant numerical models of the simulation suites we utilize in this study (Section \ref{ss:CAMELS}).
    We also discuss how we generate our synthetic spectra from the simulations (Section \ref{ss:Spectra}), how we calculate the \Lya~forest CDD and $b$-value distribution (Section \ref{ss:Lyastats}), and the different supermassive black hole (SMBH) statistics that we analyze (Section \ref{ss:BHstats}).
    In Section \ref{s:Results} we present our results, followed by a discussion in Section \ref{s:Discussion}. 
    Finally we conclude and discuss implications and future work in Section \ref{s:Conclusion}.

\section{Methodology}\label{s:Methods}
    
    \subsection{The CAMELS Simulations}\label{ss:CAMELS}

        \begin{table*}
            \centering
            \begin{tabular}{|c|c|c|c|c|}
            \multicolumn{1}{c}{ }
             & \multicolumn{2}{c}{AGN Feedback Parameters} & \multicolumn{2}{c}{Stellar Feedback Parameters} \\
             \hline Suite
             & \multicolumn{1}{c|}{\AAGN{1} Definition} & \multicolumn{1}{c|}{\AAGN{2} Definition} & \multicolumn{1}{c|}{\ASN{1} Definition} & \multicolumn{1}{c|}{\ASN{2} Definition} \\
            \hline
            Simba & Momentum flux (Eq. \ref{eqn:Simba-AAGN1}) & Jet speed (Eq. \ref{eqn:Simba-AAGN2}) & Mass loading (Eq. \ref{eqn:Simba-ASN1}) & Wind speed (Eq. \ref{eqn:Simba-ASN2}) \\
            IllustrisTNG & Energy per unit BHAR (Eq. \ref{eqn:TNG-AAGN1}) & Burstiness (Eq. \ref{eqn:TNG-AAGN2}) & Energy per unit SFR (Eq. \ref{eqn:TNG-ASN1}) & Wind speed (Eq. \ref{eqn:TNG-ASN2}) \\
            \hline \hline
             Set
             & \multicolumn{1}{c|}{\AAGN{1} Value(s)} & \multicolumn{1}{c|}{\AAGN{2} Value(s)} & \multicolumn{1}{c|}{\ASN{1} Value(s)} & \multicolumn{1}{c|}{\ASN{2} Value(s)} \\
             \hline
            1P (61)$^{\textnormal{a}}$ & [0.25 - 4] & [0.5 - 2] & [0.25 - 4] & [0.5 - 2] \\
            CV (27)$^{\textnormal{b}}$ & 1 & 1 & 1 & 1 \\
            \hline
            \end{tabular}
            \caption{Summary of CAMELS Simulation Terminology. 
            \\
            $^{\textnormal{a}}$1P Varies one parameter at a time. Variable parameters are $\Omega_m$, $\sigma_8$, \ASN{1}, \ASN{2}, \AAGN{1}, \AAGN{2}. Each suite has 61 1P simulations.
            \\
            $^{\textnormal{b}}$CV uses fiducial values for all parameters but varies the initial random seed. Each suite has 27 CV simulations.}
            \label{tab:CAMELS}
        \end{table*}
    
        The CAMELS\footnote{\url{https://www.camel-simulations.org/}} project consists of thousands of N-body and (magneto-)hydrodynamical simulations run with the AREPO, GIZMO, and GADGET codes \citep{ GADGET, camels}. 
        The CAMELS simulations are run with the same sub-grid models as pre-existing simulations while varying different cosmological and astrophysical parameters. 
        Currently the CAMELS project has made simulations run with the IllustrisTNG \citep{IllustrisTNG, Pillepich:2018a}, Simba \citep{SIMBA}, and Astrid \citep{Bird:2022} sub-grid models publicly available \citep{CAMELS-public, Ni:2023}.
        However, in this work we will focus on IllustrisTNG and Simba as the Astrid AGN feedback closely resembles that of IllustrisTNG. 
    
        For the IllustrisTNG and Simba suites, 6 parameters are varied for each sub-grid model.
        The 2 cosmological parameters explored are $\Omega_m$, the matter density, and $\sigma_8$, the variance of the linear field on 8 Mpc/$h$ scales at $z=0$.
        The 4 astrophysical parameters explored vary the different feedback models with 2 parameters assigned to stellar feedback and 2 for AGN feedback.
        The exact physical processes that the feedback parameters control vary between the different simulation sub-grid models as the feedback models differ dramatically between them.
        The extensive range of cosmological and astrophysical parameters explored makes the CAMELS simulations a unique data set to constrain astrophysical and cosmological models.  
    
        Each CAMELS simulation has $256^3$ gas resolution elements in a periodic comoving volume with a side length of 25 Mpc/$h$. 
        The mass resolution of the CAMELS simulations is similar to the original IllustrisTNG300-1 simulation. 
        Each simulation also utilizes cosmological parameters $\Omega_b=0.049$, $h=0.6711$, $n_s=0.9624$, $M_\nu = 0.0$ eV, $w=-1$, and $\Omega_K=0$. 
        In this study we do not explore simulations that vary $\Omega_m$ and $\sigma_8$, and therefore these values are set to the fiducial $\Omega_m = 0.3$ and $\sigma_8 = 0.8$ in every case.
        For each suite (i.e.\ sub-grid model for IllustrisTNG or Simba) there are 3 main ``sets'' of simulations: the 1P set (61 simulations), the CV set (27 simulations), and the LH set (1000 simulations). 
        1P stands for 1-Parameter and the simulations in this set vary one parameter at a time. 
        CV stands for Cosmic Variance and these simulations vary only the initial random seed. 
        LH stands for Latin Hypercube and these simulations vary all 6 cosmological and astrophysical parameters and the random seed simultaneously. 
        The LH set is ideal for machine learning applications. 
        A short summary of the suite and set terminology utilized in this study as well as a brief description of each feedback parameter is presented in Table \ref{tab:CAMELS}. 
    
        This study primarily focuses on the 1P set simulations that vary the stellar and AGN feedback parameters individually over the range shown in Table \ref{tab:CAMELS} while all other feedback parameters are held at their fiducial value of 1.
        Additionally, we only explore the Simba and IllustrisTNG suites as the Astrid feedback models (especially the AGN feedback model) resemble those of IllustrisTNG. 
        The feedback parameters (labeled \AAGN{1}, \AAGN{2}, \ASN{1}, and \ASN{2}) correspond to different aspects of the feedback models in IllustrisTNG vs.\ Simba as the models are dramatically different. 
        The parameters \AAGN{1} and \AAGN{2} vary aspects of the AGN feedback models while \ASN{1} and \ASN{2} vary aspects of the stellar feedback models. 
        Mathematical descriptions of the feedback parameters are provided in the following sub-sections along with an outline of the different feedback models.
    
        \subsubsection{SIMBA}\label{sss:Simba}

            \begin{figure}
                \centering
                \includegraphics[width = \linewidth, trim=0.0cm 0.0cm 0.0cm 0.0cm, clip=true]{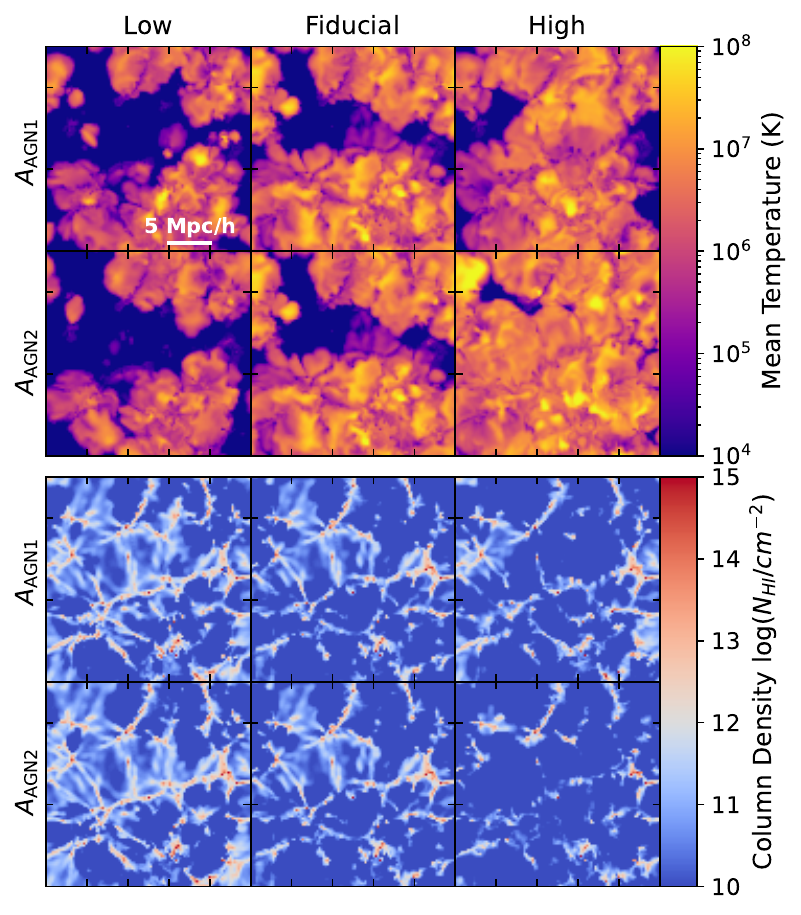}
                \caption{Simba temperature and column density projections of a single absorber slice as defined in this study (525 kpc/h, see text in \ref{ss:Lyastats} and \ref{ss:projections} for justification of this value). The top plots display the projected mean temperature weighted by mass. The bottom plots display the integrated HI column density of the slice. We show projections for varying AGN momentum flux (\AAGN{1}) and AGN jet speed (\AAGN{2}). The left plots are the lowest value for the parameter, the right plots are for the highest value, and the middle plots are the fiducial results. The projections help visualize the effect of varying the AGN feedback parameters.}
                \label{fig:Simba_proj_AAGN}
            \end{figure}

            \begin{figure}
                \centering
                \includegraphics[width = \linewidth, trim=0.0cm 0.0cm 0.0cm 0.0cm, clip=true]{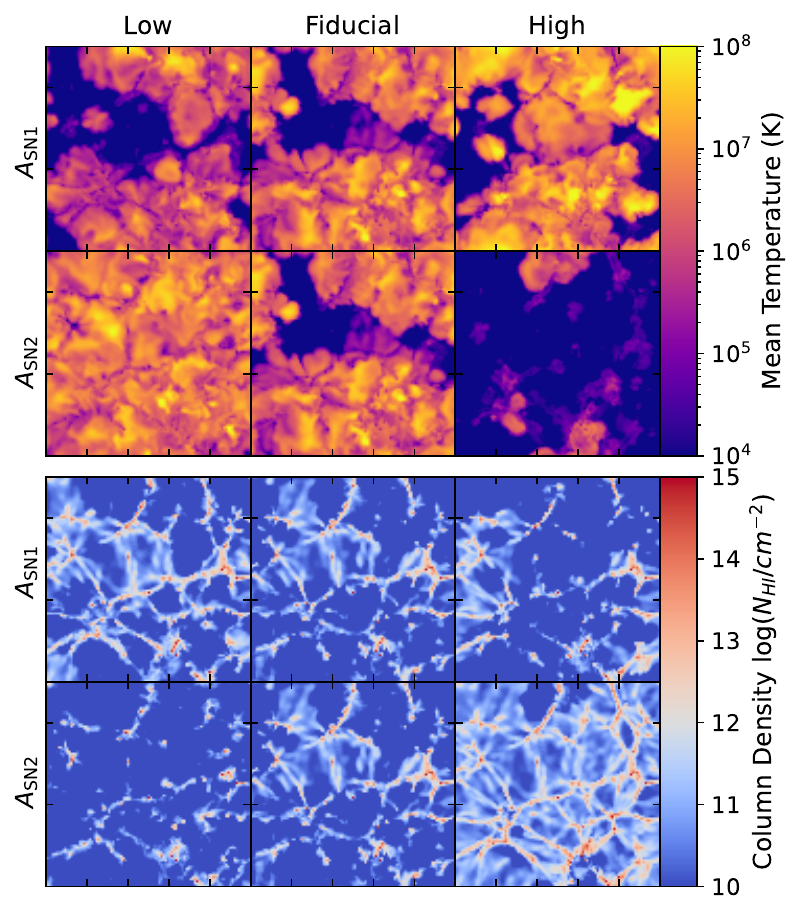}
                \caption{Same as Figure \ref{fig:Simba_proj_AAGN} except for variations of the SN mass loading factor (\ASN{1}), and SN wind speed (\ASN{2}).}
                \label{fig:Simba_proj_ASN}
            \end{figure}

            The CAMELS Simba suite follows the same sub-grid model as the original Simba simulations and is fully described in \citet{SIMBA}. 
            Simba utilizes GIZMO \citep{Hopkins:2015}, an N-body hydrodynamics code, in its Meshless Finite Mass hydrodynamics mode. 
            Radiative cooling and photoionizing heating are modeled with Grackle3.0 \citep{Smith:2017} which includes non-equilibrium evolution of primordial elements, a partially-uniform ionizing background \citep{Haardt:2012}, hydrogen self-shielding \citep{Rahmati:2013}, and metal cooling.
        
            Simba tracks chemical enrichment by Type II supernovae (SNe), Type Ia SNe, and AGB stars where eleven individual elements are tracked (H, He, C, N, O, Ne, Mg, Si, S, Ca, and Fe). 
            Dust formation in stellar ejecta, growth by accretion of metals, and destruction by thermal sputtering and SNe are also tracked \citep{Li:2019}.
    
            \subsubsection*{Simba Stellar Feedback}
    
            The stellar feedback drives galactic winds kinetically with hydrodynamically-decoupled two-phase metal-enriched winds. 30\% of those wind particles are heated based on the SNe energy and wind kinetic energy. 
            The gas elements are ejected stochastically depending on the mass loading factor $\eta_* \equiv \dot{M}_\textnormal{wind}/$SFR (Star Formation Rate) and the wind velocity $v_w$. For SIMBA, \ASN{1} controls the normalization of $\eta_*$ such that increases in \ASN{1} results in a proportional increase in the mass ejected per SFR:

            \begin{equation}\label{eqn:Simba-ASN1}
                \eta_*(M_*) = \textnormal{A}_{\textnormal{SN1}}\times
                \begin{cases}
                9\left(\frac{M_*}{M_0}\right)^{-0.317}, & \textnormal{if}\ M_* < M_0 \\
                9\left(\frac{M_*}{M_0}\right)^{-0.761}, & \textnormal{if}\ M_* > M_0
                \end{cases}
            \end{equation}

            \noindent where $M_0 = 5.2\times10^9 M_\odot$. Both the mass loading factor and the SNe wind speed prescriptions are based on the FIRE zoom-in simulations \citep{hopkins+2014, Muratov:2015, DAA:2017}.
            
            The SN wind speed at which mass is ejected is proportional to \ASN{2} plus some velocity corresponding to the potential difference between position of the SN and 0.25 times the virial radius ($R_\textnormal{vir}$). 
            This leaves \ASN{2} to control the normalization of the SN wind speed as follows
            
            \begin{equation}\label{eqn:Simba-ASN2}
                v_w = \textnormal{A}_{\textnormal{SN2}} \times 1.6\left(\frac{v_{\textnormal{circ}}}{200 \textnormal{km/s}}\right)^{0.12}v_\textnormal{circ} + \Delta v(0.25 R_\textnormal{vir}).
            \end{equation}

            In summary, the parameter \ASN{1} changes the mass loading factor of SN while
            \ASN{2} varies the SN wind speed. 
            These parameters are varied within a range of \ASN{1} $\in[0.25, 4]$ and \ASN{2} $\in[0.5, 2]$, and have a fiducial value of 1 to match the original Simba simulations.
    
            \subsubsection*{Simba Supermassive Black Hole Feedback}
    
            In the Simba suite, the black hole module is based on the gravitational torque and Bondi accretion models combined with the kinetic feedback sub-grid model in GIZMO \citep{DAA:2017a}. 
            SMBHs are seeded with mass $M_{\textnormal{seed}} = 10^4 M_\odot h^{-1}$ in galaxies with $M_* \geq 10^{9.5} M_\odot$. 
            SMBHs are re-positioned to the potential minimum of their host group if it is within $4\times R_0$. 
            $R_0$ is the size of the BH kernel which encloses the nearest 256 gas elements and any BHs within $R_0$ of each other are merged so long as their relative velocity is less than 3 times the mutual escape velocity.

            SMBH growth occurs via a two-phase model with cold gas accreted at a rate controlled through the transport of angular momentum by gravitational torques from stars \citep{Hopkins+Quataert:2011, DAA:2017a}, and hot gas is accreted via spherical Bondi accretion \citep{Bondi:1952}.
            For both modes, a radiative efficiency of 0.1 is assumed for accretion.
            The cold-gas torque-based accretion is capped at three times the Eddington limit and the hot gas Bondi accretion mode has a strict maximum of the Eddington limit.

            AGN feedback follows a two-mode model: a high Eddington ratio ($\eta \equiv \dot{M}_{BH}/\dot{M}_{\textnormal{Edd}}$ where $\dot{M}_\textnormal{Edd} = L_\textnormal{Edd}/\epsilon_r c^2$) radiative mode with high mass loading outflows, and a jet mode with faster outflows but lower mass-loading at low Eddington ratios. 
            For all AGN feedback, gas is ejected in a bipolar fashion parallel and anti-parallel to the angular momentum vector of the gas within the SMBH kernel. 
            The total momentum flux of AGN feedback follows:

            \begin{equation}\label{eqn:Simba-AAGN1}
                \dot{P}_{\textnormal{out}} \equiv \dot{M}_\textnormal{out}v_\textnormal{out} = A_\textnormal{AGN1}\times 20 L_\textnormal{bol}/c
            \end{equation}

            \noindent where $L_\textnormal{bol}=\epsilon_r\dot{M}_\textnormal{BH}c^2$ is the bolometic luminosity, $\epsilon_r=0.1$ is the radiative efficiency, and $c$ is the speed of light. 
    
            In the radiative feedback mode, the outflow velocity scales with $M_\textnormal{BH}$ like $v_\textnormal{rad} = 500 + 500(\log_{10}(M_\textnormal{BH})-6)/3$ km s$^{-1}$. 
            When the SMBH has a mass $M_\textnormal{BH} < 10^{7.5} M_\odot$ or a high Eddington ratio of $\eta > 0.2$ it produces feedback in the radiative mode. 
            Otherwise, the feedback is produced following the jet mode prescription.

            The jet mode feedback gains an additional velocity kick of $v_\textnormal{jet} = 7000 \times \textnormal{min}[1,\ \log_{10}(0.2/\eta)]$ km s$^{-1}$ such that full jet speeds are reached when $\eta < 0.02$.
            Based on both the radiative and jet feedback mode prescriptions, the outflow of any SMBH producing feedback follows:

            \begin{equation}\label{eqn:Simba-AAGN2}
                v_\textnormal{out} =
                \begin{cases}
                v_\textnormal{rad}+A_\textnormal{AGN2}\times v_\textnormal{jet} & \textnormal{if}
                \begin{cases}
                M_\textnormal{BH} > 10^{7.5}M_\odot \\
                \eta < 0.2
                \end{cases}\\
                v_\textnormal{rad} & \textnormal{otherwise}.
                \end{cases}
            \end{equation}

            \noindent In this formulation \AAGN{2} controls the maximum jet velocity achieved in the jet feedback mode. 
            At the maximum jet speed and when $M_\textnormal{gas}/M_* <0.2$ in the galaxy, SMBHs will also inject energy into the gas immediately surrounding it in the form of X-rays.
            The particles ejected in the jet mode are hydrodynamically de-coupled from the rest of the gas in the box for a time that scales with the Hubble time at the moment of ejection.
            This results in the jets traveling a distance of up to $\sim$ 10 kpc and losing a maximum of 1000 km/s in speed due to gravitational effects before re-coupling to the surroundings.
            The jet particles are heated to the virial temperature of the host halo (so long as $v_\textnormal{jet} > 2000$ km/s which is almost always the case for jet mode AGN at $z<2.0$) and are ejected in a bipolar fashion parallel to the angular momentum vector of the disk with a zero degree opening angle.

            The total momentum flux formulation in Equation \ref{eqn:Simba-AAGN1}, where $\dot{P}_\textnormal{out}$ is a set value which can be varied by \AAGN{1}, applies to both the radiative and jet modes in Simba.
            However, the way it scales with the outflow velocity ($v_\textnormal{out}$) means that significantly less mass is ejected via AGN feedback in the jet mode than the radiative mode. 
            Changing \AAGN{1} will change the amount of ejected mass in both modes but will have a greater impact on the radiative mode.
            Changing \AAGN{2} will also change the amount of mass ejected but only for the jet mode.
            For example, increasing only \AAGN{2} will cause less mass to be ejected at higher velocities in the jet mode since the momentum flux is set, but it would not affect the amount of mass ejected in the radiative mode.
            
            In summary, in the Simba suite \AAGN{1} controls the momentum flux and \AAGN{2} controls the additional jet velocity kick. 
            With how $v_\textnormal{rad}$ is calculated, it is not affected by \AAGN{1} but the amount of mass ejected ($\dot{M}_\textnormal{out}$) is.
            Conversely, the jet speed ($v_\textnormal{jet}$) is directly proportional to \AAGN{2} allowing the maximum jet boost to reach up to 14,000 km s$^{-1}$ in the most extreme case (\AAGN{2}=2). 
            The AGN feedback parameters are varied with the ranges \AAGN{1}$\in[0.25, 4]$ and \AAGN{2}$\in[0.5, 2.0]$, with fiducial values of 1.

            For additional information, see the CAMELS documentation \citep{camels} and the original Simba documentation \citep{SIMBA} and references therein.

        \subsubsection{IllustrisTNG}\label{sss:TNG}

            \begin{figure*}
                \centering
                \includegraphics[width = \linewidth, trim=0.0cm 0.0cm 0.0cm 0.0cm, clip=true]{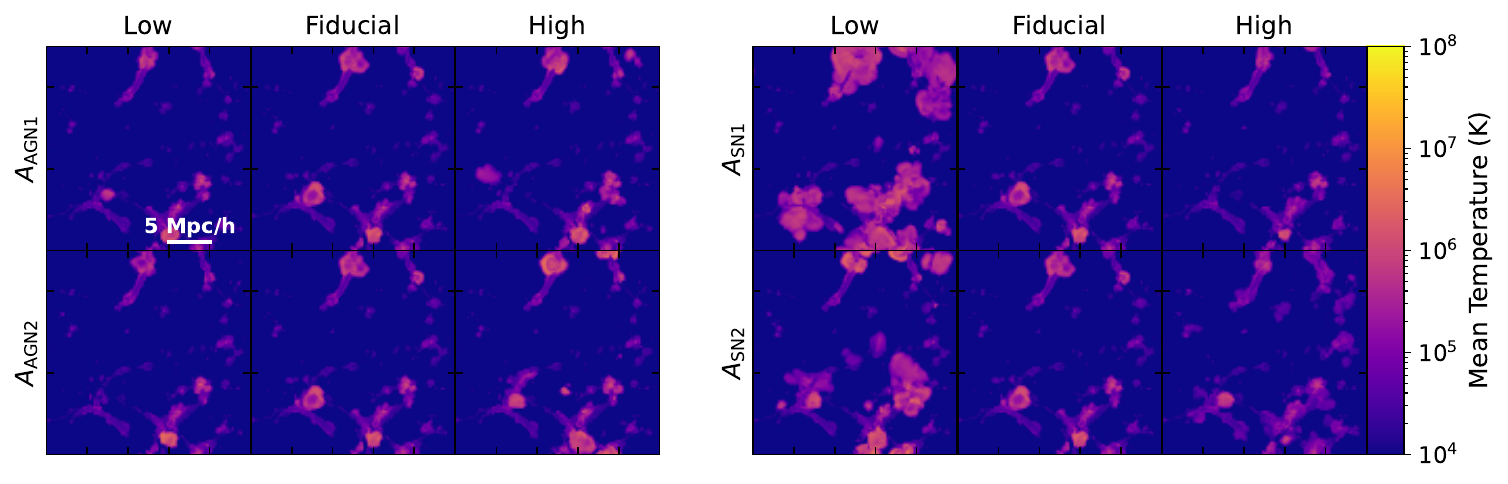}
                \caption{The same temperature projections as seen in Figures \ref{fig:Simba_proj_AAGN} and \ref{fig:Simba_proj_ASN} but for the CAMELS IllustrisTNG 1P simulation set feedback parameters. AGN feedback variations are on the left and stellar feedback variations are on the right. The parameters varied are AGN energy per SMBH accretion rate (\AAGN{1}), AGN Burstiness (\AAGN{2}), SN energy per SFR (\ASN{1}), and SN wind speed (\ASN{2}). Column density projections show minimal to no changes and are thus not included. Changes in the temperature projection due to feedback remain close to or confined to the host halos.}
                \label{fig:TNG_proj}
            \end{figure*}

            The IllustrisTNG (TNG) suite consists of magnetohydrodynamic cosmological simulations that utilize the same sub-grid models as the original TNG simulation set \citep{Pillepich:2018a}.
            The simulations are run utilizing the AREPO code \citep{Springel:2010,Weinberger:2020} and gravitational interactions evolve through the TreePM algorithm \citep{Springel:2005}.
            Radiative cooling from hydrogen and helium is implemented using the \citet{Katz:1996} network and includes cooling via line free-free emission, inverse Compton, and line cooling.
            Metals and metal-line cooling are included as described in  \citep{Vogelsberger:2012,Vogelsberger:2013}.
            TNG assumes ionization equilibrium with a \citet{Faucher-Giguere:2009} UV background and accounts for on-the-fly hydrogen column density shielding from the radiation background \citep{Rahmati:2013}.
            The star formation and interstellar medium sub-grid model is from \citet{Springel:2003}. 
            
            \subsubsection*{TNG Stellar Feedback}
    
            TNG models stellar feedback from SNe and tracks chemical enrichment from 9 elements (H, He, C, N, O, Ne, Mg, Si, Fe). 
            Enrichment is modeled from Type Ia SNe, Type II SNe, the Asymptotic Giant Branch, and neutron star-neutron star mergers.
            In TNG, stellar feedback from SNe drives kinetically implemented (with a thermal energy sub-component) galactic winds that are modeled as temporarily hydrodynamically decoupled particles. 
            These winds are stochastically and isotropically ejected from star-forming gas. 
            The total energy injected per unit star-formation rate and stellar feedback driven galactic outflow speed are modified by the parameters \ASN{1} and \ASN{2}, respectively, via the formulations below. The energy per unit SFR is:

            \begin{equation}\label{eqn:TNG-ASN1}
                \begin{aligned}
                e_w = & A_{\rm SN1} \times \Bar{e}_w\left[ f_{w,Z} + \frac{1-f_{w,Z}}{1+(Z/Z_{w,\rm ref})^{\gamma_{w,Z}}}\right]\\
                & \times N_{\rm SNII} E_{\rm SNII,51} 10^{51} \rm erg\ M_\odot ^{-1},
                \end{aligned}
            \end{equation}

            \noindent where $Z$ is the gas metallicity. 
            The wind speed of galactic outflows is:

            \begin{equation}\label{eqn:TNG-ASN2}
                v_w = A_{\rm SN2} \times {\rm max} \left[\kappa_w \sigma_{\rm DM}\left(\frac{H_0}{H(z)}\right)^{1/3},\ v_{w,\rm min}\right],
            \end{equation}

            \noindent where $\sigma_{\rm DM}$ is the local dark matter velocity dispersion and $H$ is the Hubble constant. 
            Additional parameters ($\Bar{e}_w$, $f_{w,Z}$, $Z_{w,\rm ref}$, $\gamma_{w,Z}$, $N_{\rm SNII}$, $E_{\rm SNII,51}$, $\kappa_w$, and $v_{w,\rm min}$) are constants with values and descriptions in Table 1 of \citet{Pillepich:2018a}. 
            The wind mass loading factor is then $\eta_w \equiv \dot{M}_{\rm wind} /\rm SFR = 1.8v_w^{-2}e_w \propto A_{\rm SN1}/A_{\rm SN2}^2$. 
            This results in both \ASN{1} and \ASN{2} playing a role in calculating the mass loading factor for SNe feedback.
    
            In summary, the energy injection is proportional to \ASN{1} with a fiducial value of 1 and the range of variation \ASN{1}$\in[0.25, 4]$. 
            The SN wind speed is proportional to \ASN{2} with a fiducial value of 1 and a range of variation of \ASN{2}$\in[0.5, 2]$. 
            The fiducial values correspond to the original TNG runs.
    
            \subsubsection*{TNG Supermassive Black Hole Feedback}
    
            In the TNG suite, SMBH particles are seeded in halos with mass $M_\textnormal{halo} > 5\times10^{10}M_\odot h^{-1}$ with mass $M_\textnormal{seed} = 8 \times 10^5 M_\odot h^{-1}$. 
            BH accretion in TNG uses the \citet{Bondi:1952} accretion prescription capped at the Eddington limit.
            A radiative efficiency of 0.2 is assumed for all BH growth.

            The SMBH feedback in TNG includes three modes: thermal, kinetic, and radiative. 
            The radiative feedback mode is always on, adding the radiation flux of the SMBH to the cosmic ionizing background heating the gas in and around the host halo. 
            In addition to the radiative mode, either the kinetic or thermal mode is active. 
            Which mode the SMBH produces feedback in depends on the Eddington ratio. 
            The efficiency fraction (fraction of accreted mass converted into energy) for the thermal mode is a constant $0.02$.
            The efficiency fraction in the kinetic mode is calculated as min$[0.2,\rho/(0.05\rho_{\rm SFthresh}]$), where $\rho$ is the density of the gas around the SMBH and $\rho_{\rm SFthresh}$ is the star formation threshold density.
            
            In both the kinetic and thermal feedback modes, energy is injected in the `feedback sphere' of the SMBH. 
            The feedback sphere has a size that scales with resolution $\propto m_{baryon}^{-1/3}$. 
            The size of the sphere is roughly constant within a simulation but varies slightly with the particles neighboring the SMBH \citep{IllustrisTNG, Pillepich:2018a}.
            Unlike the stellar feedback, there is no de-coupling of the cells within the feedback sphere from hydrodynamical forces or radiative cooling. 
            The transition from thermal to kinetic feedback mode happens at Eddington ratios ($\eta$) lower than some threshold value ($\chi$) with
    
            \begin{equation}
                \eta \equiv \frac{\dot{M}_{BH}}{\dot{M}_{\textnormal{Edd}}} \leq \chi,\ \ \chi = \textnormal{min}\left[0.002 \left(\frac{M_{BH}}{10^8M_\odot}\right)^2, 0.1 \right].
            \end{equation}
    
            \noindent Due to this definition, the transition from high-accretion thermal feedback mode to low-accretion kinetic feedback mode occurs at $M_{BH} \sim 10^8M_\odot$. 
            In CAMELS, the high-accretion rate thermal mode is set as an injection of thermal energy into the defined sphere around the SMBH. 
            The efficiency fraction of the mass-to-energy conversion for energy injection is $\dot{E}_\textnormal{high} = 0.02 \times \dot{M}_{BH} c^2$ where $\dot{M}_{BH}$ is the instantaneous SMBH accretion rate and $c$ is the speed of light.
            
            The CAMELS simulations explores variations of the low-accretion rate kinetic mode. 
            For the kinetic mode, energy is accumulated until a certain threshold is reached (since the last event) after which the energy is injected into the feedback sphere in a random direction (averaging over multiple events the injections become isotropic).
            The energy threshold at which injection occurs is:
    
            \begin{equation}\label{eqn:TNG-AAGN2}
                E_{\textnormal{inj,min}} = A_\textnormal{AGN2} \times 10\sigma_\textnormal{DM}^2 m_\textnormal{enc}
            \end{equation}
    
            \noindent where $\sigma_\textnormal{DM}^2$ is a one-dimensional dark matter velocity dispersion around the SMBH and $m_\textnormal{enc}$ is the gas mass in the feedback sphere. 
            The amount of energy produced per accretion event is proportional to both \AAGN{1} and the gas density around the SMBH ($\rho$) with:
    
            \begin{equation}\label{eqn:TNG-AAGN1}
                \dot{E}_\textnormal{low} = A_\textnormal{AGN1} \times \textnormal{min}\left[ \frac{\rho}{0.05\rho_\textnormal{SFthresh}}, 0.2\right] \dot{M}_{BH} c^2
            \end{equation}
    
            \noindent where $\rho_\textnormal{SFthresh}$ is the density threshold for star-formation. 
            
            Given these relationships, when \AAGN{1} is doubled so is the efficiency at which mass is converted into energy per accretion event.
            When \AAGN{2} is doubled so is the energy threshold in which injection happens, therefore larger values for \AAGN{2} results in less frequent but stronger AGN feedback events. 
            In this sense, for the TNG SMBH feedback model in CAMELS, \AAGN{1} controls the amount of energy produced per SMBH accretion event while \AAGN{2} controls the burstiness and strength of the SMBH feedback. 
    
            The parameters controlling AGN feedback in the CAMELS TNG suite are intertwined such that increasing \AAGN{1} alone does not increase the strength of any individual AGN feedback injection but only increases the frequency at which these events occurs. 
            If \AAGN{2} is also increased, the feedback events grow stronger but at the expense of the frequency at which the events occur. 
            The range explored for these parameters are \AAGN{1} $\in$ [0.25, 4] and \AAGN{2} $\in$ [0.5, 2.0] with a fiducial value of 1 corresponding to the original TNG runs.

            For additional information, see the CAMELS documentation \citep{camels} and the original TNG documentation \citep{Pillepich:2018a} and references therein.

    \subsection{Synthetic Spectra}\label{ss:Spectra}

        We generate synthetic spectra from the CAMELS simulations using the publicly available \textit{fake-spectra}\footnote{\url{https://github.com/sbird/fake_spectra}} code outlined in \citet{Bird:2015, Bird:2017} with MPI support from \citet{Qezlou:2022}. 
        From simulation snapshots the code generates and analyzes mock spectra. The \textit{fake-spectra} package is fast, parallel, and is written in C++ and Python 3 with the user interface being Python-based. 
     
        Column densities (CDs, $N_{\rm HI}$) are computed by interpolating the neutral hydrogen mass in each gas element to the sightline using an SPH (smoothed particle hydrodynamics) kernel. 
        The method used is based on what is appropriate for the corresponding simulation; for the CAMELS-TNG simulations a tophat (or uniform) kernel is used while for CAMELS-Simba a cubic spline kernel is used. 
        The CDs have units of neutral hydrogen atoms (HI) per cm$^{-2}$.

        We generate 5,000 sightlines randomly placed in each simulation box; a number found to be sufficient for avoiding variations due to sampling \citep{Tillman:2023}. 
        We do not add noise to the spectra generated from the simulation box.
        Adding noise to the spectra will not affect the results of this study as it would largely alter the lowest CD values, where observational errors are large, and the $b$-values predicted, which we do not compare to observations herein.
        In this study, we focus primarily on CDs of $ 10^{12} < N_{\rm HI} < 10^{15}$ cm$^{-2}$. Below $10^{12}\ \textnormal{cm}^{-2}$ the lines are too faint to detect and characterize at current observational sensitivities, while above $10^{15}$ cm$^{-2}$ lines are saturated and it becomes difficult to accurately determine CDs from flux. 
        Additionally, absorbers at the highest CDs are rarer, which makes them difficult to study both observationally, and in the small-box (25 Mpc h$^{-1}$)$^3$ CAMELS simulations.
        \citet{Tillman:2023} found that at least for TNG, the CAMELS boxes produce a converged CDD as compared to the original TNG100-1 simulation, within observational error bars. 

     \subsection[Lyman-alpha Statistics]{\Lya~Statistics}\label{ss:Lyastats}
     
        The CCD ($f(N_{\rm HI})$) is defined as 
     
         \begin{equation}
            f(N_{\rm HI}) = \frac{d^2N}{d\log (N_{\rm HI}) dz} = \frac{F(N_{\rm HI})}{\Delta \log N_{\rm HI} \Delta z}  
         \end{equation}
     
         \noindent where $F(N_{\rm HI})$ is the fraction of absorbers with column densities in the range [$N_{\rm HI}$, $N_{\rm HI}+\Delta N_{\rm HI}$], and $\Delta z$ is the redshift distance of the sightline. 
         The CDD describes the number of absorbers within a logarithmic column density bin width and redshift distance. 
     
         To calculate the CDD we utilize a direct integration method as described in \citet{Tillman:2023}. The particles in 525 kpc/h slices are amalgamated into absorbers that are then used to calculate column densities for the CDD. 
         As discussed in \citet{Tillman:2023} this method and the defined size of the absorbers results in a well converged CDD. Variations in the chosen absorber size (reasonable sizes ranging from 300 kpc/h to 800 kpc/h) results in a less than $\sim 15$\% effect on the CDD at CD values lower than $10^{12.5}$ cm$^{-2}$, a less than 1\% effect at higher CD values, and all differences are well within 1$\sigma$ of observational error bars. 
         It has been previously found that the direct integration method as compared to Voigt profile fitting provides a CDD that is converged within 1 sigma with respect to observational errors \citep{Gurvich2017}.
         At the lowest column densities ($\sim 10^{12}$ cm$^{-2}$) the CDD produced via Voigt profile fitting diverges from the direct integration significantly likely due to the difficulty in fitting those absorbers with an automated procedure. 
         However, the divergence is still within 1 sigma of the observational error bars where observational data exists and does not influence the main results of this paper.

         To calculate the $b$-value (Doppler width) distribution, Voigt profile fitting is necessary. 
         To conduct our fits we utilize the Voigt fitting algorithm included in the \textit{fake-spectra} package\footnote{\url{https://github.com/sbird/fake_spectra/blob/master/fake_spectra/voigtfit.py}}. 
         The algorithm is closely based on that of AUTOVP \citep{AutoVP}.
         Peaks are found, fit, and iteratively removed after which the peaks are refit to the spectrum all together.
         The algorithm starts with the largest peaks and continues until adding another peak to the fit no longer improves the fit by some chosen significance value. 
         Another stopping condition is when the largest peak remaining is less than $10^{-4} \times \textnormal{max}[1, \textnormal{max}(\tau)]$ where $\tau$ is optical depth. 
         These conditions are in place since, as the fit continues, smaller and smaller peaks are more likely to become fitting errors rather than an actual absorber. 
         It is due to this that automatic Voigt profile fitting is so difficult, and this is why the lower end of the CDD diverges when using the Voigt fitting method. 
     
         The original Voigt profile fitting algorithm in the \textit{fake-spectra} package minimizes the squared difference between the fit and the original spectrum using the Nelder-Mead algorithm. 
         For our work we use a modified version of the Voigt profile fitter that minimizes the earth mover's distance using a simple bounds limited-memory Broyden–Fletcher–Goldfarb–Shanno algorithm.
         We found that this method had an easier time fitting smaller peaks in the spectra and ran into fewer errors during computation, but this choice did not change the overall results of our study.

         The $b$-value is affected by both the temperature and turbulence of an absorber. This manifests mathematically as:
    
         \begin{equation}\label{eqn:b-value}
             b \propto \left(\frac{2kT}{m_H} + \sigma_v^2\right)^{1/2}
         \end{equation}

         \noindent where $\lambda_0$ is the wavelength of the \Lya~transition, $m_H$ is the mass of a hydrogen atom, $k$ is the Boltzman constant, $T$ is the temperature, and $\sigma_v$ is the root mean-square turbulent velocities of the absorber.
         The $b$-value can also have components from peculiar and Hubble velocities.
         At least at high redshift ($z\sim 2$ to 4) the $b$-value of absorbers tends to be dominated by the Hubble flow (corresponding to the physical width of the absorber along the line of sight) but is also affected by thermal broadening and pressure support along the Jean's length, but the narrowest absorption features are dominated by thermal broadening \citep{Bryan:2000,Peeples:2010I}. At lower redshifts, such as those explore in this work, the Hubble flow becomes less important in absorber broadening.

         The formulation above also puts a lower limit on $b$-values we examine in this work. 
         Assuming an absorber is non-turbulent and recognizing that the cooling floor of the simulations is $T_\textnormal{min}=10^4$ K, we should assume a minimum $b$-value of $b_\textnormal{min}\approx 13$ km s$^{-1}$. 
         We limit the upper bounds of our $b$-value analysis to 100 km s$^{-1}$ as this is where observational data is available. However, this bound does not affect the results of our study because there are so few absorbers at such high values.

         We do not compare to observational values or the original simulations since the $b$-value distribution is more sensitive to mass resolution than the CDD \citep[see Appendix in][]{Burkhart_2022}. 
         Lower numerical resolution causes a nonphysical broadening of the $b$-value distribution shifting to higher $b$-values, making the comparison to observations challenging \citep[artificial broadening due to numerical resolution limits as seen in][]{Peeples:2010I}.
         Due to this, the $b$-value distributions generated from the CAMELS simulations are most robustly interpreted when comparing between simulations using the same numerical resolution, as done in this paper. See Appendix \ref{Appendix:CVandConverge} for additional discussion on \Lya~forest statistics convergence.

         We explored introducing a Gaussian line spread function (LSF) with a full width half maximum of 6.5 to our data to determine how noise due to the HST-COS instrumentation might affect our results. While this is not the exact LSF utilized when fitting COS data, it represents a good approximation and should be sufficient in determining the general instrumentation effects on this study. The LSF broadens the absorption features, which leads to larger $b$-values and smaller CDs. As previously mentioned, noise in the spectra largely affects lower $b$ and CD values. The effects of introducing this LSF on the $b$-value and CD distributions are small and our main results remain unchanged therefore the results presented herein include no instrumentation or random noise.
         
         For both the CDD and the $b$-value distributions, we will explore how these statistics vary for the various feedback parameters explored in CAMELS. 
         To help visualize the effects on the various statistics we also generate projection plots for the various CAMELS simulations of both absorber CDs and average temperatures. 
         The CD projections help interpret the CDD while the temperature projections can help interpret both the CDD and the $b$-value distribution. 
         These projections can be seen in Figures \ref{fig:Simba_proj_AAGN}, \ref{fig:Simba_proj_ASN} and \ref{fig:TNG_proj}. 

    \subsection{Black Hole Statistics}\label{ss:BHstats}

        We also analyze certain BH statistics to determine the extent to which stellar feedback can suppress SMBH growth and thus AGN feedback, and to see how efficient AGN feedback is at self-suppressing SMBH growth. 
        We analyze the accretion rate density for SMBHs in the CAMELS simulations which is the total accretion rate from all SMBHs in the box divided by the box volume ($\dot{\rho}_{BH} = \sum_{i}^{} \dot{M}_{BH,i} / 25$ Mpc$^3$ for all SMBHs $i$).
        The energy emitted in feedback scales with accretion rate, thus $\dot{\rho}_{BH}$ gives a sense of the total energy in the box due to SMBH feedback.
 
        For CAMELS-TNG, the thermal AGN mode produces feedback as thermal energy injected into the area immediately surrounding the BH whereas the kinetic mode drives matter and energy out to larger distances. 
        Considering this, we expect the kinetic mode to be more likely to affect the \Lya~forest as opposed to the thermal mode in TNG. 
        However previous work has shown, by completely removing the kinetic feedback, that this feedback mode does not affect the CDD in any meaningful way \citep{Tillman:2023}. 
        In this work, we instead find the radiative AGN feedback mode to be the most impactful on the forest statistics in TNG.
        The contribution from the radiative mode scales with the amount of matter being accreted, therefore, for TNG, the contribution to the UVB from AGN feedback scales with $\dot{\rho}_{BH}$.
    
        For Simba, the jet mode feedback velocity boost depends on $\eta$ such that a lower $\eta$ produces faster jet speeds up to a maximum jet speed boost of \AAGN{2}$\times$7000 km s$^{-1}$ for $\eta \leq 0.02$. 
        As seen in \citet{Christiansen:2020, Tillman:2023}, the jet mode in Simba can have a dramatic effect on various IGM statistics particularly due to the ability for jets to reach far into the IGM.
        Analyzing the AGN jet mode through the SMBH accretion rate will aid in interpreting the effects of varying the feedback parameters.
        Lower values for $\dot{M}_{BH}$ in the jet mode correspond to higher jet speeds but less energetic events overall.
        
        For both TNG and Simba, we supplement the accretion rate statistics with the number of SMBHs in the different feedback modes. 
        We will analyze these BH statistics for variations of the CAMELS-TNG and CAMELS-Simba feedback parameters \ASN{1}, \ASN{2}, \AAGN{1}, and \AAGN{2}.

\section{Results}\label{s:Results}

    In this study we explore statistics from the \Lya~forest and from the SMBHs themselves to determine what influences the feedback has on the neutral hydrogen in the IGM and why those effects manifest. 
    We analyze the mass-weighted mean temperature and CD projections in Section \ref{ss:projections}. 
    We also explore the CDD and $b$-value posterior distribution function in Sections \ref{ss:CDD} and \ref{ss:bvalue} respectively. 
    Finally, we look at SMBH statistics regarding the Eddington ratio, accretion rate, number of SMBHs in the simulation, and the number of SMBHs accreting in each feedback mode in Section \ref{ss:BHstatsresults}. 

    \subsection{Temperature and Column Density Projections}\label{ss:projections}

        The projection plots in Figures \ref{fig:Simba_proj_AAGN}, \ref{fig:Simba_proj_ASN} and \ref{fig:TNG_proj} clearly show how varying the different feedback modes in TNG and Simba affects the IGM. 
        The projections show a 525 kpc/h slice of the simulations corresponding with the typical size of an absorber found in the \Lya~forest as defined in this study. 
        For most definitions, used in previous works, the absorber length is found to be less than 500 kpc/h \citep{Tonnesen:2017} but for temperatures as low as $10^4$ K an absorber length of 800 kpc/h could be expected \citep{Peeples:2010}.
        However, for the analysis methods used herein, it was found that variations of the absorber length within these ranges had less than a 15\% impact on the resulting CDD \citep{Tillman:2023}.
        The projection plots are good visual indicators of the overall effect of varying the stellar and AGN feedback parameters in CAMELS.
    
        Figures \ref{fig:Simba_proj_AAGN} and \ref{fig:Simba_proj_ASN} show mass-weighted mean temperature projections and column density projections for the Simba suite when varying the feedback parameters.
        Figure~\ref{fig:Simba_proj_AAGN} shows projections for variations in the AGN feedback parameters (\AAGN{1} and \AAGN{2} respectively), and Figure~\ref{fig:Simba_proj_ASN} shows variations in the stellar feedback parameters (\ASN{1} and \ASN{2} respectively).
        For both figures, the top two rows display temperature projections and the bottom two rows are column density projections.
        The left column is for the lowest value of the parameter explored, the middle column indicates fiducial results, and the right column displays the highest value, as indicated at the top of each figure.

        Varying AGN momentum flux (\AAGN{1}) and AGN jet speed (\AAGN{2}) have clear consequences for both the temperature in the box and the amount of neutral hydrogen in the IGM. 
        With increases in either AGN parameter value, we see higher temperatures and less neutral hydrogen populating the IGM. 
        However, much of the change in HI abundances is seen in CDs outside of the range of interest for the \Lya~forest.
        Changes in \AAGN{2} demonstrate these effects more dramatically than does \AAGN{1}. 
    
        Varying the SN mass loading factor (\ASN{1}) and the SN wind speed (\ASN{2}) in Simba also has a clear impact on both the temperature and amount of neutral hydrogen in the box.
        For increases in the SN mass loading factor we see increases in temperature and decreases in the amount of neutral hydrogen.
        The extent of this effect appears similar to that of the AGN feedback parameters. 
        However, increasing the SN wind speed has the opposite effect, with higher wind speed decreasing the temperature and increasing the fraction of neutral hydrogen. 
    
        The SN wind speed has the most dramatic effect on the IGM out of the four feedback parameters explored in CAMELS-Simba. 
        Stellar feedback as implemented in these simulations does not, by itself, cause IGM scale effects.
        If heating from SN directly influenced the forest, then increases in SN wind speed should see a decrease in \Lya~absorbers, however the opposite is true.
        As we will explore in more detail in this section and the discussion (Section \ref{s:Discussion}), the stellar feedback's effect on the IGM comes indirectly through its impact on SMBH growth and AGN feedback.
        
        The complex interaction between stellar feedback and SMBH growth in simulations has been explored in many studies \citep{Booth:2013,Dubois:2015,Habouzit:2017,DAA:2017SMBHgrowth,vanDaalen:2019,Lapiner:2021,Tillman:2022,Byrne:2023,Delgado:2023}.
        Star formation is able to regulate the growth of galaxies and their central SMBHs, and stronger stellar feedback can both reduce the need for AGN feedback to regulate star formation as well as suppress SMBH growth. Weaker AGN feedback will then diminish the overall impact on the IGM.
        We explore SMBH statistics for variations in the feedback parameters to help illustrate this interplay between AGN and stellar feedback.

        For the TNG suite we only display temperature projections because the column density projections exhibit no visible changes when varying feedback parameter values.
        Figure \ref{fig:TNG_proj} resembles closely that of Figures \ref{fig:Simba_proj_AAGN} and \ref{fig:Simba_proj_ASN}, but shows only the temperature projections for CAMELS-TNG. 
        AGN feedback parameter variations are on the left and stellar feedback variations are on the right.
        The temperature of the IGM in TNG is overall much lower than that in Simba.
        Increases in the AGN feedback parameters show minimal increases in temperature and those changes appear largely confined to host halos and the surrounding area.
        Decreases in the stellar feedback parameters show more substantial temperature changes but the effects still reside within $\sim$1-2 Mpc of host halos. 
        Similar to the Simba suite, the stellar feedback parameter variations in TNG also highlight the importance of AGN-stellar feedback interplay.
        However, the impacts are not as pronounced as they are for the Simba suite as the TNG AGN feedback model has minimal effect on the IGM.

    \subsection{Column Density Distribution Functions}\label{ss:CDD}

        \begin{figure*}
            \centering
            \includegraphics[width = \linewidth, trim=0.0cm 0.0cm 0.0cm 0.0cm, clip=true]{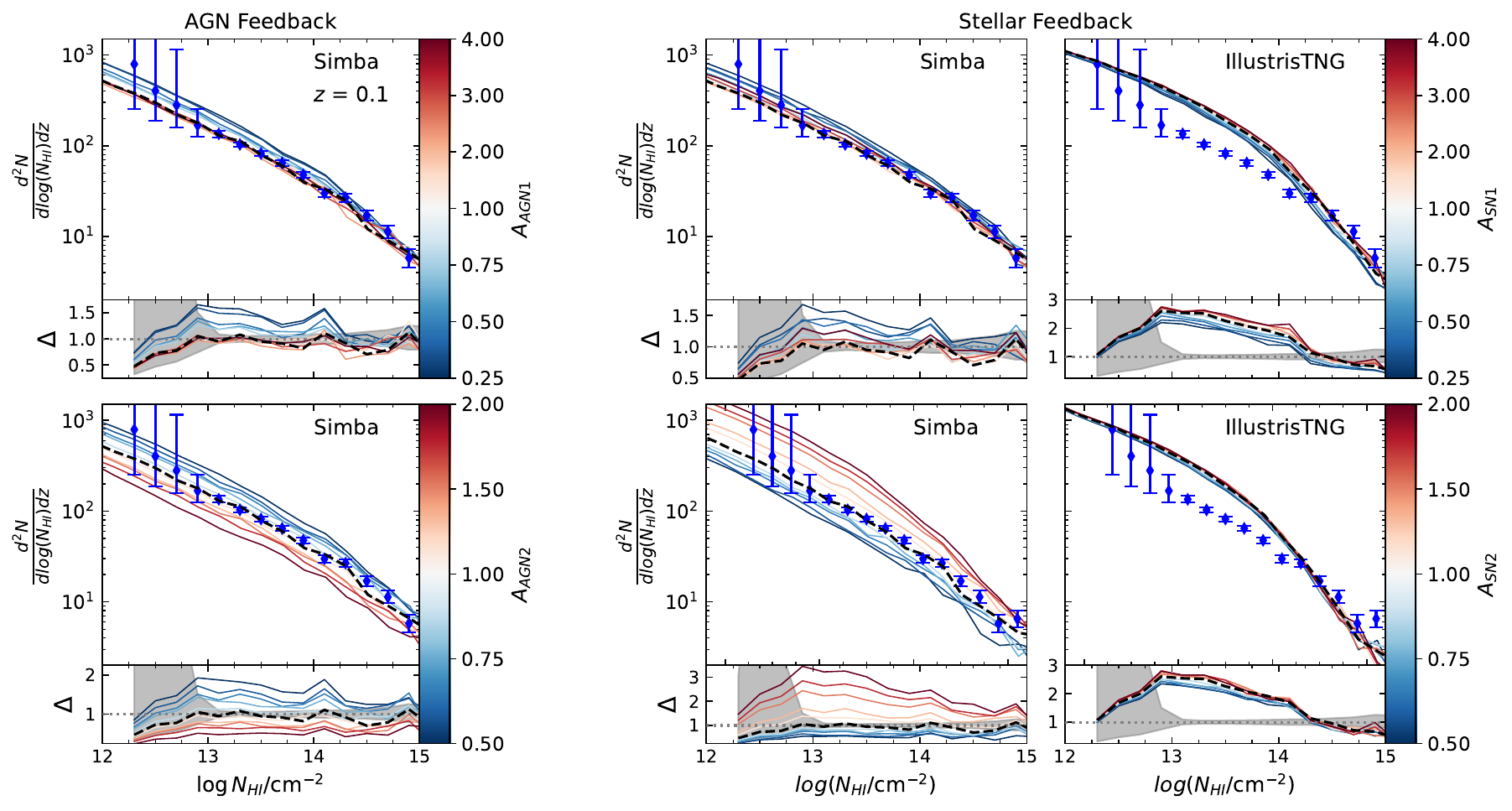}
            \caption{The Simba and TNG CDDs at $z=0.1$ for the CAMELS 1P set varying the AGN feedback parameters (left two plots) and stellar feedback parameters (right four plots). The blue scatter points are observational data from D16. The blue lines represent a decrease in the parameter value (labeled in corresponding color-bars) while the red lines represent an increase. The dashed black lines are the fiducial results. The panel below each plot displays the ratio, $\Delta$, of the parameter variation result to the observed D16 data. The gray shaded region corresponds to the observational error bars. The TNG results when varying the AGN feedback parameters are not shown as no changes in the CDD are seen.}
            \label{fig:CDD}
        \end{figure*}

        Next, we examine the CDD to quantify the impacts of the parameter variations seen in the previous section. 
        The two leftmost plots of Figure~\ref{fig:CDD} show the CDD for different values of \AAGN{1} and \AAGN{2}. 
        Only the Simba suite is displayed in these plots as the TNG results for AGN feedback showed no discernible differences, as evident in the projection plots.
        
        For Simba, as implied in the projection plots, decreases in the AGN feedback parameters show increases in the amount of neutral hydrogen in the \Lya~forest.
        However, increases in \AAGN{1}, relative to the fiducial value, show no observable effect on the CDD (with respect to the observational error bars) while increases in \AAGN{2} result in less neutral hydrogen overall.
        Increases in the \AAGN{1} parameter affect the temperature and HI fraction of gas not associated with the \Lya~forest since differences are clearly visible in the projection plots but not in the CDDs. 
        The gas affected is too hot and at too low column densities to be associated with the forest.

        The right four plots of Figure \ref{fig:CDD} show the same results for the stellar feedback parameters. 
        For the Simba suite, decreasing \ASN{1} shows an increase in neutral hydrogen absorbers as implied by the results in Figure~\ref{fig:Simba_proj_ASN}. However, increasing \ASN{2} shows an increase in HI absorbers.
        For the remainder of the parameters (\ASN{2} for Simba, and \ASN{1} and \ASN{2} for TNG) larger values result in more absorption while smaller values result in less. 
        The effect is the most dramatic for the SN wind speed in Simba and minimal for the stellar feedback parameters in TNG. 
        However, both of these results imply the suppression of SMBH growth by stellar feedback.
        We discuss the relationship between stellar and AGN feedback further in Section~\ref{ss:BHstatsresults}.

        Figure~\ref{fig:CDD_zevo} shows the redshift evolution of the TNG and Simba suites from $z=2.0$ to $z=0$. 
        In these plots, the TNG CDD is corrected to utilize the \citet{Haardt:2012} UV ionizing background to match the one used in Simba. 
        Removing the difference in the assumed UVB model allows for easier comparison of AGN and stellar feedback effects between the two simulations. 
        In TNG, the AGN radiative feedback mode adds to the assumed ionizing background which effectively results in a slightly stronger UVB than the assumed \citet{Faucher-Giguere:2009} UVB.
        Since in Figure~\ref{fig:CDD_zevo} we only account for the difference between the \citet{Haardt:2012} and \citet{Faucher-Giguere:2009} UVB models, the magnitude of the difference made by the radiative feedback mode in TNG is well visualized at $z=2.0$.
        If we instead rescale both of the simulations at $z=2.0$ to have the same \Lya~mean flux, the predicted CDDs of TNG and Simba lie on top of one another.
        At $z=2$ the shape of the CDD is nearly identical between the two simulations, after which they diverge around $z\sim 0.5$.

    \subsection{The Doppler Width ($b$-value)}\label{ss:bvalue}

\begin{figure*}
            \centering
            \includegraphics[width = \linewidth, trim=0.0cm 0.0cm 0.0cm 0.0cm, clip=true]{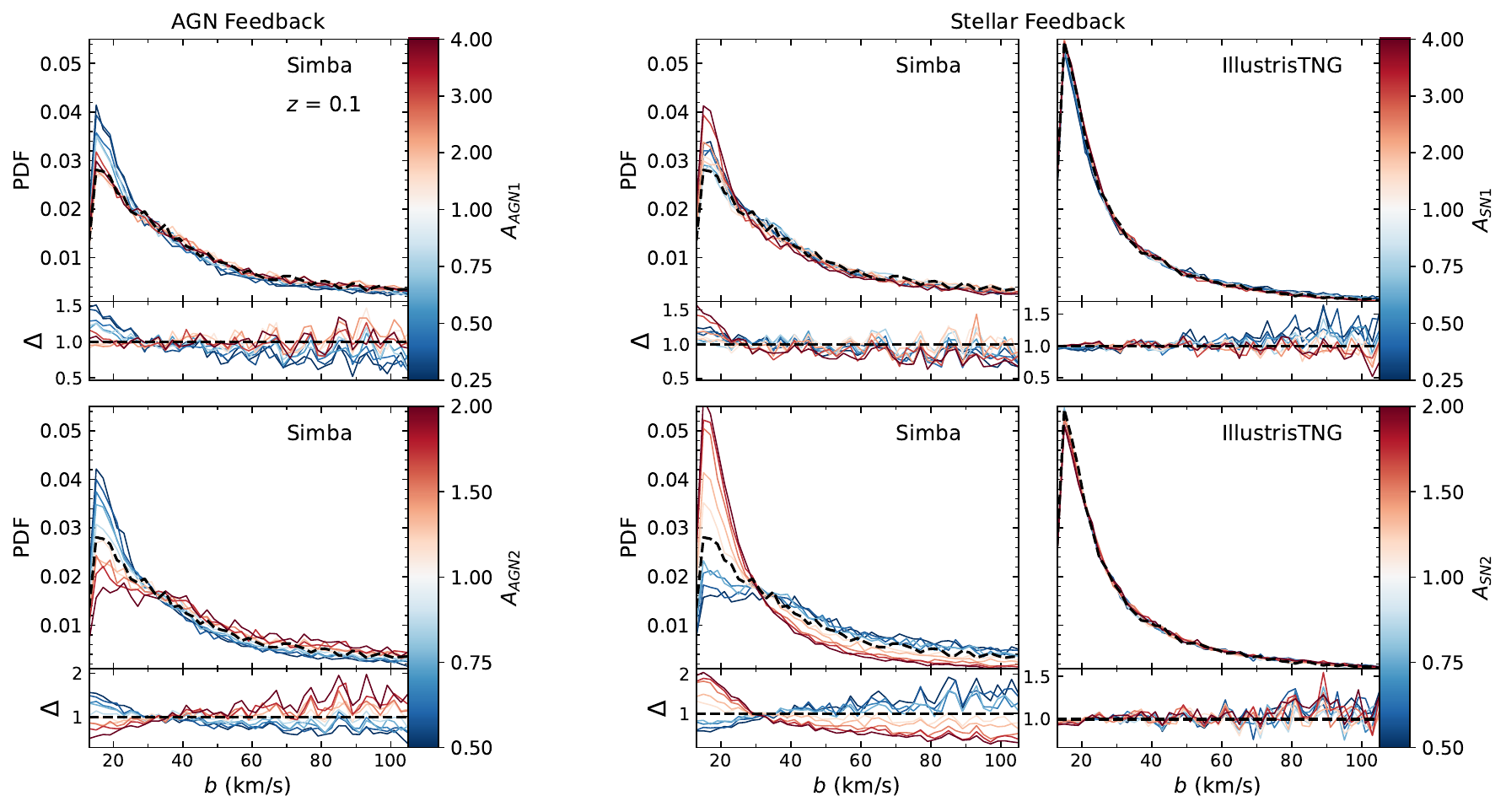}
            \caption{Same as Figure \ref{fig:CDD} but for the $b$-value distribution and $\Delta$ is the ratio of the parameter variation results to the fiducial results. The distribution is normalized to integrate to unity. No change is seen in the b-value distribution for the TNG simulations when varying AGN feedback so those results are excluded.}
            \label{fig:bvalue}
        \end{figure*}

        Similar to the CDD analysis, we explore the $b$-value distributions for variations in the CAMELS feedback parameters for both the TNG and Simba suites. 
        Figure~\ref{fig:bvalue} shows how the $b$-value probability density function (PDF) varies with the feedback parameters. 
        Overall Simba has a broader distribution of $b$-values due to having a hotter IGM than TNG.
        However, the peak of the distribution is around 20 km/s for both simulations.

        The left two plots of Figure \ref{fig:bvalue} show the effects of AGN feedback in Simba on the $b$-value distribution.
        Higher $b$-values are observed for stronger AGN feedback, with the AGN jet speed (\AAGN{2}) demonstrating a stronger effect than the mass loading factor (\AAGN{1}).
        Increases in the AGN feedback parameters, in Simba, show an overall increase in the IGM temperature (seen in Figure~\ref{fig:Simba_proj_AAGN}) so increased $b$-values, seen in Figure~\ref{fig:bvalue}, are expected. 
        As with the CDD results, we do not include the $b$-value distributions for variation in the TNG AGN feedback parameters as no discernible difference is seen.

        The right four plots of Figure \ref{fig:bvalue} show variations in the stellar feedback parameters for both Simba and TNG.
        For Simba, decreases in the stellar feedback parameters show a shift to higher $b$-values with the SN wind speed (\ASN{2}) having a more dramatic effect than the SN mass loading factor (\ASN{1}). 
        For TNG, variations in the stellar feedback parameters have only a small effect with marginal shifts to higher $b$-values for lower SN energy per unit SFR (\ASN{1}) and higher SN wind speed (\ASN{2}) but these differences are subtle and would not be measurable in observations. 
        
        The results from Figure \ref{fig:bvalue} parallel the effects seen on the CDD, but perhaps to a smaller degree than was seen for the CDD. 
        This can be explained by the fact that both the CDD and the $b$-value distribution have a temperature scaling, although the $b$-value tends to have a stronger temperature correlation. These temperature effects are due to the AGN jets in Simba and the radiative mode in TNG.

    \subsection{Supermassive Black Hole Statistics}\label{ss:BHstatsresults}

        Now that we have demonstrated the impact of varying CAMELS parameters on the \Lya~forest statistics, we explore its cause.
        When varying the different feedback parameters available in the CAMELS simulation suite, Simba's feedback exhibited clear effects on the \Lya~forest CDD and the $b$-value PDF, while TNG's feedback produced no effects that could be observationally measured, at least for the parameter variations explored. 
        These results are consistent with previous work comparing the AGN feedback models in Simba and TNG and their effects on the \Lya~forest CDD \citep{Tillman:2023}.
        
\begin{figure*}
            \centering
            \includegraphics[width = \linewidth, trim=0.0cm 0.0cm 0.0cm 0.0cm, clip=true]{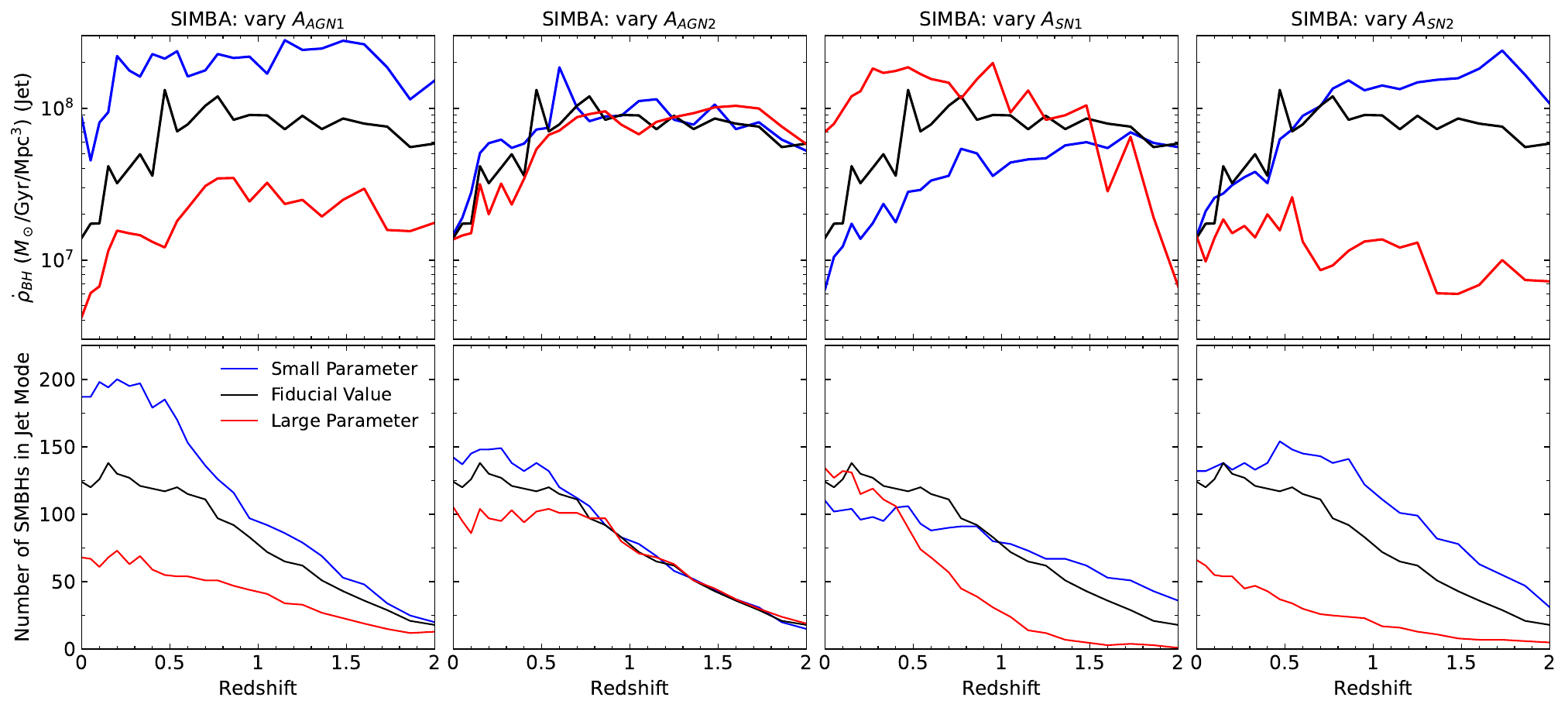}
            \caption{BH accretion rate density in the jet mode and overall number of SMBHs in the jet mode when varying different feedback parameters in the CAMELS-Simba simulations. \textbf{All plots:} The black lines correspond to fiducial results, the blue lines correspond to the smallest value for the parameter varied, and the red lines correspond to the largest value for the parameter varied. \textbf{Top row:} The SMBH accretion rate density $\dot{\rho}_{BH}$ in the AGN jet feedback mode vs. redshift for different values of the parameters. \textbf{Bottom row:} The number of SMBHs producing jet mode feedback for parameter variations.}
            \label{fig:Simba_SMBH_stats}
        \end{figure*}

\begin{figure}
            \centering
            \includegraphics[width = \linewidth, trim=0.0cm 0.0cm 0.0cm 0.0cm, clip=true]{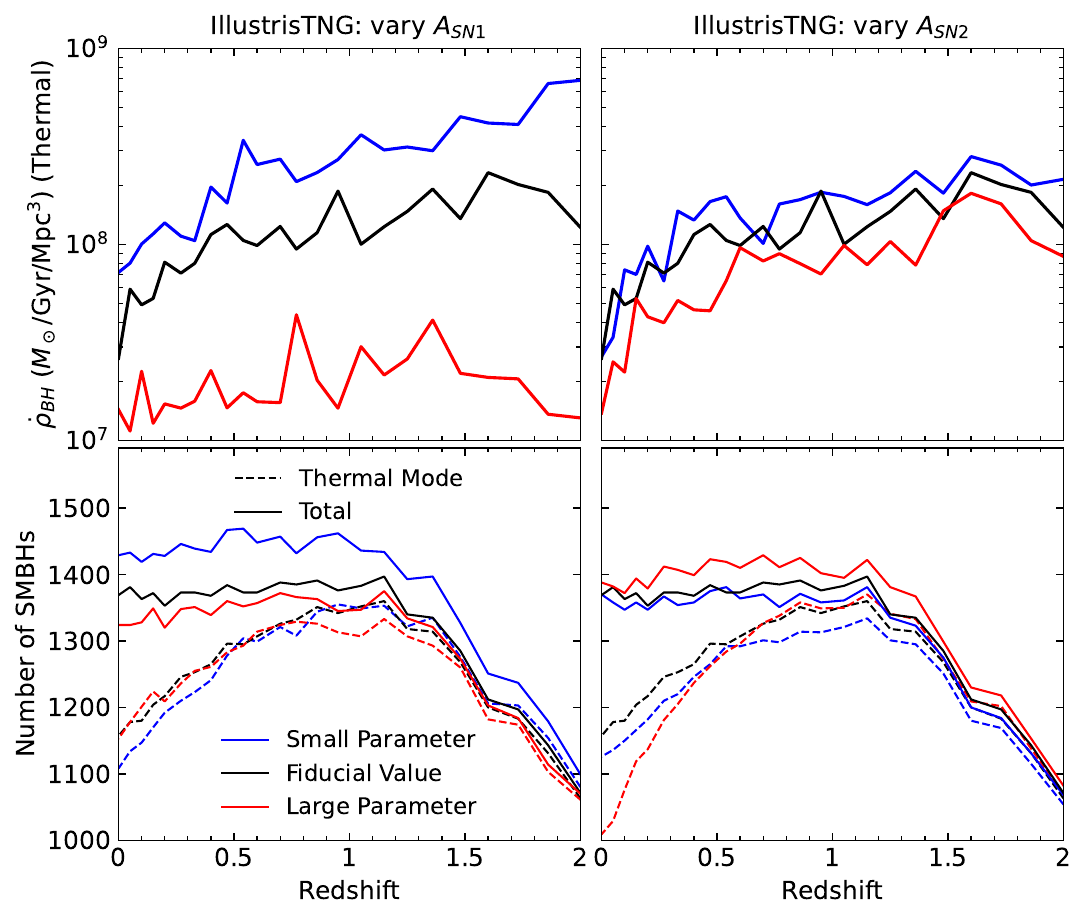}
            \caption{Similar to Figure \ref{fig:Simba_SMBH_stats} but instead for the CAMELS-TNG simulations. The top row display the SMBH accretion rate density for the thermal feedback mode. The bottom row displays the total number of SMBHs (solid lines) along with the number of SMBHs in the thermal mode (dashed lines). The results for variations in the AGN feedback parameters are not shown since those statistics are largely unaffected.}
            \label{fig:TNG_SMBH_stats}
        \end{figure}

        However, the stellar feedback parameters show a subtle effect on the CDD and $b$-value distribution for TNG. 
        This may be initially surprising, as we do not expect the effects of stellar feedback to directly manifest in low CD \Lya~forest absorbers, which can be greater than a Mpc away from any SNe. The scale of a single SN (which individually drives winds on the scale of pcs) is not relevant when thinking about the larger scale impacts of galactic winds contributed by multiple SNe. However, we find in this work, for both Simba and TNG, that changes in \Lya~statistics due to varying stellar feedback are not due to galactic wind impact but rather due to stellar feedback suppressing SMBH growth. 
        Therefore, these results imply that the AGN feedback in TNG has at least a small impact on the \Lya~forest and that the parameters chosen to be varied for TNG's AGN feedback are not representative of said impact.
        To explore the interplay between stellar and AGN feedback, we analyze SMBH statistics over the redshift range of $z=2$ to 0.
        The results are displayed in Figures \ref{fig:Simba_SMBH_stats} and \ref{fig:TNG_SMBH_stats}.
        Exploring how SMBH accretion and seeding in the simulations vary with stellar feedback will help illustrate how SMBHs are affected.
        
        The top row of Figure \ref{fig:Simba_SMBH_stats} shows the accretion rate density $\dot{\rho}_{BH}$ for the Simba AGN jet feedback mode. 
        Note that all of these SMBHs have masses $M_{BH} > 10^{7.5} M_\odot$ and Eddington ratios $\eta \leq 0.2$ in order to be in the jet mode.
        The second row shows the number of SMBHs producing jet mode feedback. 
        Each column of plots represents a different feedback parameter that is being varied (labeled at the top), with the red lines corresponding to the highest values for the feedback parameters, the black lines corresponding to the fiducial results, and the blue lines correspond to the lowest values. 

        Figure \ref{fig:TNG_SMBH_stats} is similar to Figure~\ref{fig:Simba_SMBH_stats} in that it shows $\dot{\rho}_{BH}$ for the TNG thermal feedback mode in the top two row and the number of SMBHs in the bottom row. 
        The radiative mode is always on and scales with the overall accretion rate but the kinetic mode is radiatively inefficient so we are mostly interested in the thermal mode \citep{IllustrisTNG}.
        Variations in $\dot{\rho}_{BH}$ due to the AGN parameter variations are not shown since no effect is seen.
    
        Figures \ref{fig:TNG_SMBH_stats} and \ref{fig:Simba_SMBH_stats} clearly show that the stellar feedback parameters in both simulation suites have an impact on SMBH growth in the simulations. 
        We discuss the interplay of stellar and AGN feedback for each simulation suite individually.

        \subsubsection{Simba SMBH Statistics}\label{sss:SimbaBHstats}

        \subsubsection*{Simba AGN feedback parameters}

            By focusing on the first row of Figure \ref{fig:Simba_SMBH_stats}, we find that when increasing \AAGN{1}, the BHs in the box have overall larger AGN jet velocities (via Equation \ref{eqn:Simba-AAGN2} i.e.\ smaller $\dot{M}_{BH}$ means smaller $\eta$) starting at higher redshifts. 
            However, these BHs are less massive overall and are accreting less mass \citep[similar to what was found in][when increasing AGN momentum flux]{DAA:2017a}.
            These results imply that BH fueling is hindered by increases in \AAGN{1}. 
            This form of self-regulation from the BH feedback results in a decrease in energy propagated out to IGM scales.
            Additionally, a higher value for \AAGN{1} results in fewer AGN producing feedback in the jet mode (bottom panel).
            Despite AGN jets reaching maximum velocities starting at higher redshifts and AGN ejecting more mass in feedback events, the fact that less AGN jet feedback occurs with overall less energy available (lower $\dot{\rho}_{BH}$) results in no impact on the CDD.
            
            Decreasing the AGN momentum flux increases the number of BHs producing feedback in the jet mode, increases the overall mass of these BHs, and increases the amount of mass these BHs are accreting. 
            However, the $\dot{M}_{BH}$ values of these BHs tend to be larger at higher redshifts which results in a delay of when maximum jet speeds are reached, and the lower value for \AAGN{1} means less mass is ejected from the accreting SMBH overall. 
            Therefore, less energy overall is being propagated far out to IGM scales.
            The greater amount of energy available (larger $\dot{\rho}_{BH}$) is not able to overcome this fact resulting in a null effect.

            The effect of varying the AGN jet speed (\AAGN{2}) is more intuitive, as $\dot{\rho}_{BH}$ does not vary significantly. 
            The energy per feedback event does not change for variations in \AAGN{2} but the fraction of energy going into the outflow velocity does.
            This allows for jets to travel further before hydrodynamically re-coupling to the gas in the box and depositing kinetic and thermal energy. 
            Increasing jet speed also decreases the number of BHs producing jet feedback at low redshifts. 
            This is likely due to faster jet speeds suppressing stellar growth in halos which prevents SMBHs from being seeded and from growing. 
            However, reducing the number of SMBHs seeded this way does not reduce the impact that the AGN jets have on the \Lya~statistics. 

            \subsubsection*{Simba stellar feedback parameters}
            
            Figure \ref{fig:Simba_SMBH_stats} shows that increasing the mass loading, from the fiducial value, increases $\dot{\rho}_{BH}$ of the SMBHs, at $z<1.0$, in the jet mode while simultaneously decreasing the number of SMBHs producing feedback in the jet mode, at $z \gtrsim 0.5$.
            The suppression of AGN jet feedback at higher redshifts, due to increases in the SN mass loading factor hinders the ability of the jet feedback to remove HI from the \Lya~forest by $z=0.1$.
            This effect is most likely dominated by the slower jet speeds from higher $\dot{M}_{BH}$ rather than the smaller number of SMBHs considering results from other parameter variations.
            
            Mass loading factors smaller than the fiducial value results in lower $\dot{\rho}_{BH}$ and fewer SMBHs producing jet feedback by $z=0.1$. 
            Despite AGN jet feedback in the box reaching maximum jet velocities at higher redshifts, it appears the effect of an overall reduced SMBH population producing jets at $z<0.8$ dominates.
            Moreover, lower $\dot{\rho}_{BH}$ means less energy from feedback is in the box.
            However, the increase in HI absorbers due to larger \ASN{1} is not as dramatic as the increase due to smaller \ASN{1}.
               
            More intuitive than the mass loading parameter is the SN wind speed parameter (\ASN{2}). 
            Values for $\dot{\rho}_{BH}$ remain quite similar by $z=0$ but the number of SMBHs producing jet feedback decreases with increasing SN wind speed and $\dot{\rho}_{BH}$ is dramatically lower at higher $z$. 
            Due to the large decrease in the number of SMBHs producing jet feedback and the amount of energy from feedback when increasing SN wind speed, the overall HI surviving in the IGM increases and the temperature of that HI tends to be cooler.
            Strong stellar feedback in the form of faster SN wind speeds efficiently suppresses SMBH growth and feedback in the simulation box.
            Since the BH seeding mechanism in Simba is tied to the stellar mass of the galaxy, it is clear the lack of SMBHs results from suppressed stellar growth within galaxies due to the fast SN wind speeds.

    \subsubsection{TNG SMBH Statistics}\label{sss:TNGBHstats}
            
        As previously shown in \citet{Tillman:2023}, the kinetic mode in the TNG AGN feedback model has no discernible effect on the low-$z$ \Lya~forest.
        From Figure \ref{fig:TNG_proj} a subtle heating effect can be seen when varying the AGN feedback parameters explored in CAMELS and more prominent heating is seen for variations in stellar feedback. 
        From here, it is clear the feedback parameters affect the temperature in the box to some degree, but on Mpc scales this is small. 
        Despite there being twice as many SMBHs in TNG as in Simba no effect is seen for the \Lya~forest statistics explored herein when varying the CAMELS-TNG AGN feedback parameters.
        No changes were seen in the SMBH statistics when varying the AGN feedback parameters, either, implying that the AGN feedback parameters in CAMELS-TNG play no role in self-suppression of SMBH growth. 
        This makes sense as both \AAGN{1} and \AAGN{2} modify the kinetic feedback mode whereas the thermal feedback mode self-regulates the SMBH growth the most \citep{IllustrisTNG}.
        
        Increasing the stellar feedback parameters tends to decrease the value of $\dot{\rho}_{BH}$ and in the case of \ASN{1} (energy per unit SFR) the number of BHs in the simulation is decreased as well. 
        These results imply that the AGN feedback in TNG affects the forest to some degree since small variations in the \Lya~statistics are seen when stronger stellar feedback suppresses BH growth.
        As discussed previously, the radiative feedback mode is the AGN feedback mode that affects the IGM in TNG. 

        The radiative mode adds flux directly to the cosmic ionizing background and heats the gas around the host halo. 
        The flux and heat added will scale with the accretion rate of the SMBHs in the box and both of these effects have implications for the gas in the IGM. 
        The increased energy in the ionizing background will be most impactful on the CDD.
        In the forest, CD is proportional to the inverse of the photoionizing rate but for temperature the relation is weaker, following $N_\textnormal{HI} \propto T^{-0.7}$. 
        We not only see the temperature scaling affecting the CDD but we also see a slight difference in the $b$-value distributions since $b$ scales with temperature as in Equation~\ref{eqn:b-value}.

        The AGN thermal feedback mode is more radiatively efficient so changes to $\dot{\rho}_{BH}$ from the thermal mode are expected to have the most impact on \Lya~statistics. 
        From Figure \ref{fig:TNG_SMBH_stats}, increases in both \ASN{1} and \ASN{2} relative to the fiducial values show decreases in $\dot{\rho}_{BH}$ (but less so for \ASN{2}).
        This results in less flux added to the UVB and thus more HI absorption in the forest explaining the shift in the TNG CDD seen in Figure~\ref{fig:CDD}.
        There is also a reduction in the number of SMBHs in total for increased \ASN{1} and a reduction in thermal mode SMBHs for increased \ASN{2}.
        These changes to the number of SMBHs seem to have minimal impact most likely due to the change being relatively small when compared to the total number in the box (note the y-axis in the bottom panels only extends from 1000 to 1600).
        
        On the other hand, decreases in \ASN{1}, relative to the fiducial value, show an increase in the total number of SMBHs and in $\dot{\rho}_{BH}$. 
        This should result in more flux being added to the UVB and thus reduce the number of absorbers in the forest, which is exactly what is seen from the CDD.
        Decreases in \ASN{2} result in a small increase in $\dot{\rho}_{BH}$ and a small decrease in the number of SMBHs in the box.
        Larger $\dot{\rho}_{BH}$ should result in a stronger UVB and thus less HI absorption but since these changes in $\dot{\rho}_{BH}$ for \ASN{2} are less dramatic than for \ASN{1} the effect will be less discernible. 
        The changes to the CDD due to smaller \ASN{2} are much smaller than changes due to smaller \ASN{1} which is exactly as expected.

        Since the AGN feedback parameters explored thus far have shown no impact on the \Lya~forest statistics we also analyze results from additional parameters explored by CAMELS. 
        The CAMELS IllustrisTNG extended 1P set is similar to the original 1P set but explores 22 additional parameters. Variations of all these parameters simultaneously composes the SB28 set \citep[the original 6 parameters plus 22 more results in 28 total parameters][]{Ni:2023}.
        The additional SMBH sub-grid model parameters explored in this set are: The Bondi rate multiplier, the high-accretion mode feedback efficiency, the Eddington rate multiplier for the BH accretion rate limiter, the BH seed mass, the BH radiative efficiency (fraction of rest mass energy released in feedback), the Eddington ratio for the transition between the AGN feedback modes, and the steepness of the mass transition between the AGN feedback modes. 
        As with the original AGN feedback parameters explored herein, we found no observable difference, relative to the observational error bars, in the \Lya~forest CDD for independent variations of any of the parameters listed above.
        At least, no difference was found that showed more dramatic results than the stellar feedback parameters for TNG. 
        It appears that apart from the radiative mode in TNG, the TNG AGN feedback model has essentially no effect on the low-z \Lya~forest. 

\section{Discussion}\label{s:Discussion}

    In the last section we found that the galactic feedback in Simba, both from AGN and SNe, can have a dramatic impact on the predicted \Lya~forest. 
    AGN momentum flux, SN mass loading, and SN wind speed all have the ability to regulate the growth of SMBHs and variations in these parameters can manifest in different predictions for the \Lya~forest CDD and $b$-value distribution -- specifically, feedback that lowers the number of SMBHs or decreases the AGN feedback flux will result in more absorption. 
    The AGN jet speed in Simba has the largest direct effect on the predicted \Lya~forest statistics explored herein, with faster jets heating a larger swath of the IGM and decreasing the number of absorbers.
    Indeed, at late times, in fiducial Simba, the AGN heating from jets becomes almost volume filling and the low CD absorbers are particularly vulnerable while the higher CD absorbers are less susceptible to strong shocks on re-coupling \citep{Christiansen:2020, Tillman:2023}.
    This results in an impact on the CDD preferentially at the low CD end, flattening the distributions, as seen in Figure~\ref{fig:CDD}.
    
    While the kinetic and thermal AGN feedback in TNG appears not to have an impact on the predicted \Lya~forest, it is clear that the AGN radiative feedback mode has at least a small effect. 
    Variations in the stellar feedback parameters in TNG, that can regulate the growth of SMBHs in the box, help illustrate the consequences that AGN have on the predicted \Lya~forest in TNG.
    
    Multiple previous studies have posed the use of the low-$z$ \Lya~forest as a tool for constraining galactic feedback models and the results herein further motivate that idea.
    How to move forward with this idea relies on a careful analysis of other processes that have an effect on the predicted \Lya~forest not limited to but including the assumed UVB.
    We also must acknowledge the limitations of analyzing large-scale parameter spaces such as the CAMELS simulations.
    We discuss these ideas and more in the following sub-sections.

    \subsection{Degeneracy in \Lya~Forest Statistics}\label{ss:Degens}

    \begin{figure*}
            \centering
            \includegraphics[width = \linewidth, trim=0.0cm 0.0cm 0.0cm 0.0cm, clip=true]{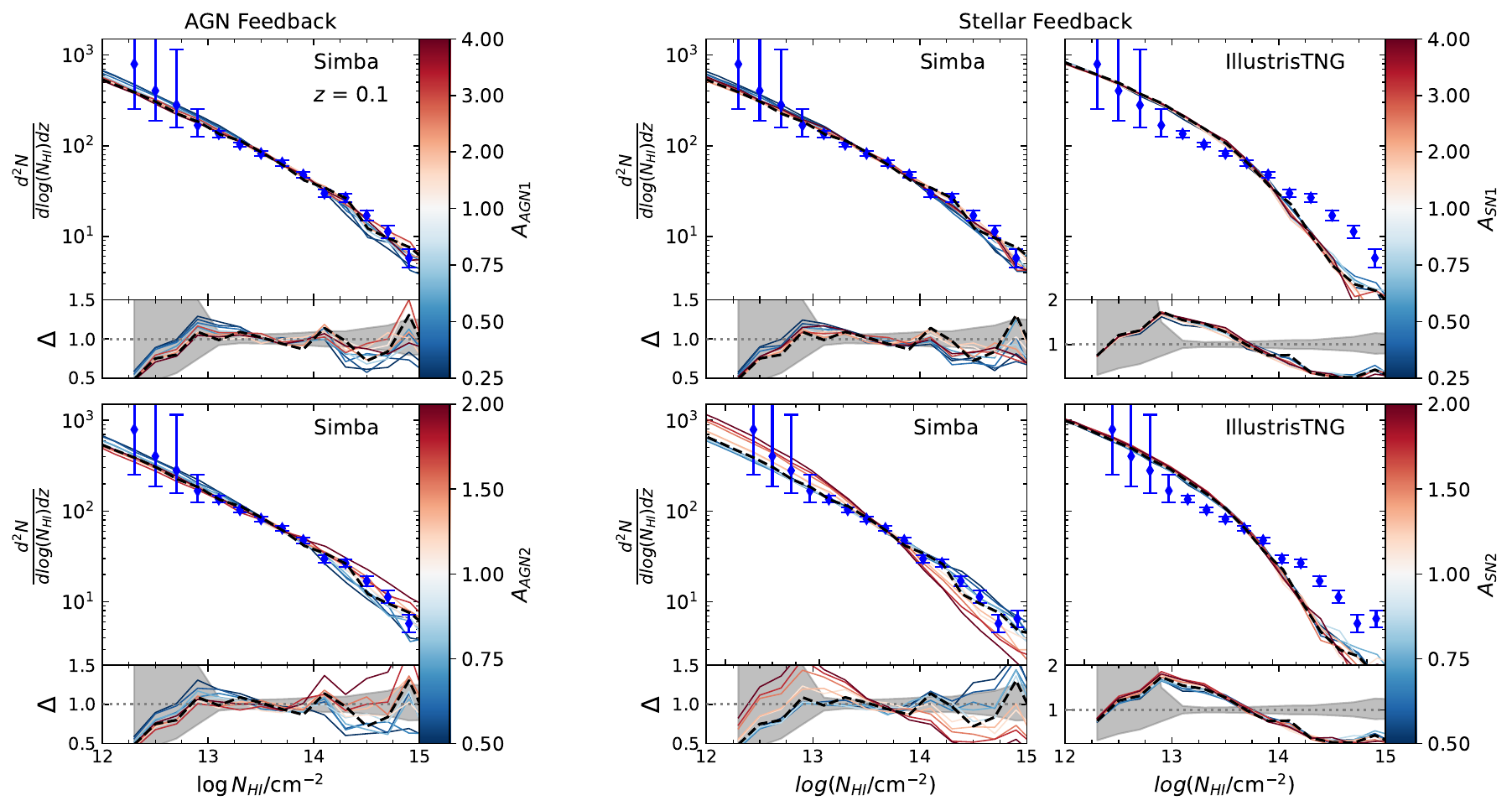}
            \caption{Same as Figure \ref{fig:CDD} but fitting the results of each parameter variation to the observational data by allowing the strength of the UVB to vary. A table including the UVB correction value and reduced $\chi^2$ values for the fit can be found in Table \ref{tab:UVBcorr} in Appendix \ref{Appendix:BestFits}.}
            \label{fig:CDD_UVBcorr}
        \end{figure*}
            
        In \citet{Tillman:2023}, the degeneracy between the UVB and AGN feedback models was analyzed. 
        That study found that the impact of AGN jet feedback on the CDD resulted in both a simple translation of the distribution function that was degenerate with changes in the UVB, as well as a change in the slope of the CDD that could not be replicated with a change in the UVB.
        With regards to scaling the UVB, assuming that the \Lya~forest is in ionizing equilibrium and optically thin, column density scales with the photoionization rate of hydrogen as $N_{HI} \propto \Gamma_{HI}^{-1}$. 
        This results in a normalization shift as changing the value of $\Gamma_{HI}$ for the UVB applies to all the gas in the simulation box, and the effects are largely equivalent for gas in CDs associated with the forest. 
        Temperature changes also affect the CDD due to the aforementioned assumptions leading to the relationship $N_{HI} \propto T^{-0.7}$.
        Since the heating due to AGN jets is able to reach distances on the order of Mpcs into the diffuse IGM, low density gas is efficiently heated.
        This means certain CDs of the forest are affected more than others causing a slope change in the CDD. This is the result seen in \citet{Tillman:2023} where AGN jet feedback in Simba flattens the CDD.
        
        Understanding other impacts on the forest has important implications for how one might constrain AGN feedback in the future. 
        For example, a better fit to observational data might be achieved with a weaker AGN feedback model accompanied by a slightly stronger UVB.
        This scenario could be likely in Simba since the fiducial AGN jet mode aggressively blows out gas which can negatively affect group statistics \citep{Robson:2020,Robson:2021, Oppenheimer:2021,Lovisari:2021,Yang:2022}. But as we saw when varying the jet speed, the amount of gas ejected is less important than the ability for that ejected gas to transport heat out to the diffuse IGM.

        We also recognize an additional important factor with a degenerate effect on \Lya~forest statistics in Simba.
        The stellar feedback parameters, especially the SN wind speed, can be efficient at suppressing supermassive black hole growth and thus affect the forest. 
        For Simba, increases in the SN wind speed amount to an effect similar to that of decreasing the AGN jet speed. 
        In fact, smaller changes in SN wind speed are required for the same change in the CDD or $b$-value distribution when varying AGN jet speed.
        Instead of decreasing the AGN jet speed, one can slightly increase SN wind speed and achieve the same result.

        \begin{figure*}
            \centering
            \includegraphics[width = \linewidth, trim=0.0cm 0.0cm 0.0cm 0.0cm, clip=true]{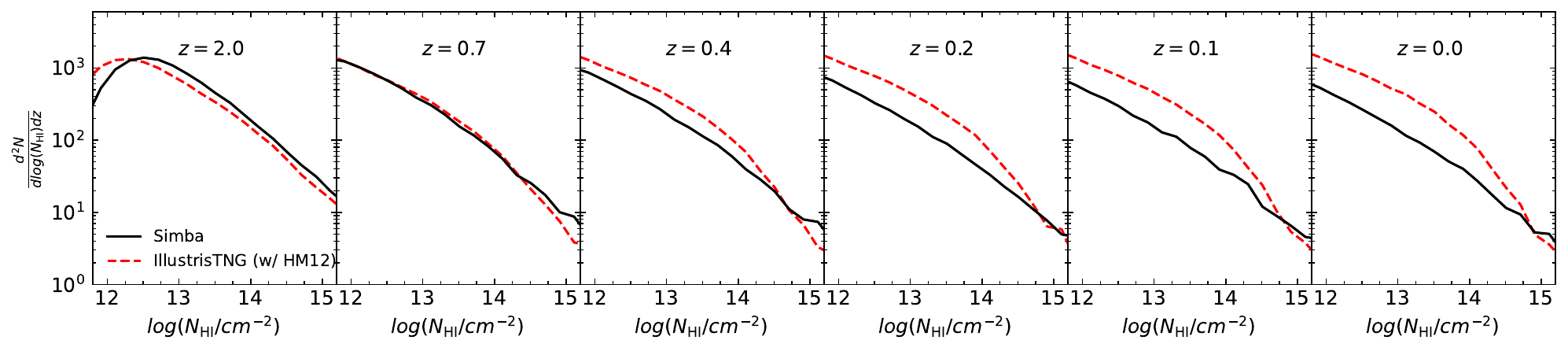}
            \caption{The redshift evolution of the SIMBA (black solid line) and TNG (red dashed lines) simulation column density distributions. The TNG CDD has been adjusted so as to use the same \citet{Haardt:2012} UVB model as used in Simba. This has been done to more accurately determine what differences are caused by either feedback or large-scale environment as opposed to the ionizing background.}
            \label{fig:CDD_zevo}
        \end{figure*}

        \begin{figure}
            \centering
            \includegraphics[width = \linewidth, trim=0.0cm 0.0cm 0.0cm 0.0cm, clip=true]{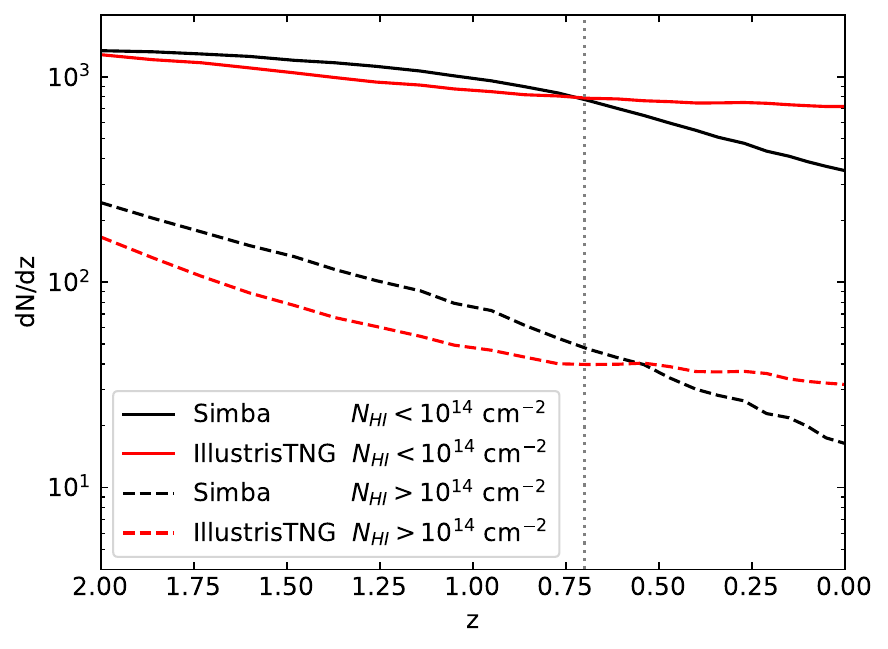}
            \caption{The redshift evolution of the total number of absorbers per redshift distance for CD bins of $10^{12}<N_{HI}<10^{14}$ cm$^{-2}$ and $10^{14}<N_{HI}<10^{17}$ cm$^{-2}$. Results from CAMELS-Simba and CAMELS-TNG are shown. The dotted grey line marks $z=0.7$, the threshold at which the shape of the CDDs diverge significantly between Simba and TNG. Like in Figure \ref{fig:CDD_zevo}, the CAMELS-TNG data is corrected in post-processing to utilize the \citet{Haardt:2012} UVB for comparison to the CAMELS-Simba data.}
            \label{fig:dNdz}
        \end{figure}

    \subsection{Redshift effects}\label{ss:zevo}
        
        Figure \ref{fig:CDD_zevo} shows the CDD redshift evolution of TNG vs.\ Simba. 
        In the plots, the TNG results have been corrected to use the \citet{Haardt:2012} UVB (the same UVB as in Simba) to make the results more directly comparable. 
        We check if the difference between TNG and Simba at $z=2.0$ could stem from a different mean flux since the simulations not only have different UVBs but TNG also includes a radiative AGN feedback mode which adds flux to the assumed UVB.
        Rescaling to the observed effective optical depth, as seen in \citet{Kim:2007}, results in CDDs at $z=2.0$ that are converged below $N_\textnormal{HI} < 10^{14}$ cm$^{-2}$, but for $10^{14} < N_\textnormal{HI} < 10^{15}$ cm$^{-2}$ Simba produces slightly more absorbers on the order of 0.1 dex.

        The CDDs of Simba vs TNG at $z=2.0$ are almost identical in that they produce similar shapes and abundances.
        As the universe evolves to lower redshift the differences between the simulations reveal themselves. 
        The slope of the Simba CDD remains largely unchanged while a steepening in the TNG CDD appears above $N_\textnormal{HI} = 10^{14} ~{\rm cm^{-2}}$ becoming more dramatic at lower $z$. 
        The differences seen between $z=2$ and $z=0$ could be partially attributed to the density of gas the forest probes at these redshifts and the extent to which feedback might affect those densities. 
        An absorber of CD $N_\textnormal{HI} = 10^{14}~{\rm cm^{-2}}$ at $z=2$ is likely to be probing higher density gas than the same CD absorber at $z=0$ \citep{dave:1999}. 
        If we instead looked at similar density gas in the forest at $z=2$ we may see a larger difference than we do in Figure \ref{fig:CDD_zevo}.

        To further explore the difference in the shapes of the evolving CDDs, we analyze the total number of absorbers per redshift distance $dN/dz$ in two different CD bins in Figure \ref{fig:dNdz}. The $10^{12}<N_{HI}<10^{14}$ cm$^{-2}$ CD range probes weak absorbers while the $10^{12}<N_{HI}<10^{14}$ cm$^{-2}$ range probes strong absorbers. From $z=2$ to $z\sim0.7$ the abundance of weak absorbers is about the same between Simba and TNG, but, as seen in their CDDs, the two simulations diverge at lower redshift and that divergence is mainly driven by the loss of absorbers in Simba. AGN heating in Simba likely causes the drop in absorber abundance seen while in TNG the number of absorbers in each bin appears to plateau after $z\sim0.7$.
        
        The difference between the CDDs predicted by Simba and TNG as the universe evolves highlights additional degenerate effects not explored herein.
        The differences are not due to cosmic variance or cosmology as the CAMELS 1P set simulations use identical initial conditions and cosmological parameters.
        In \citet{Tillman:2023} removing the AGN jet feedback in Simba did not produce a slope change dramatic enough to match the TNG results.
        However, the simulations explored in \citet{Tillman:2023} did not have the same box size, initial conditions, and had slightly different cosmological values.
        Regardless, those results imply that part of the difference seen between the two simulations (starting at $z\lesssim 0.7$) might arise from differences in the temperature and density distribution of the IGM in Simba and TNG that are caused by some factor in addition to AGN jet feedback.
        A closer look at individual absorbers in the simulations via a cross-correlation method is likely necessary to fully explain these differences.

    \subsection{The effect of cosmic variance.}\label{ss:CV}
    
        Figure \ref{fig:CDD_CV} shows the CDD for Simba and TNG when allowing the initial random seed of the simulation to vary.
        The maximum variation of the CDD in the CV set due to initial conditions appears within $\sim$2-2.5 sigma (relative to the smallest observational error bars) and the interquartile range of variation is well within 1 sigma.
        For TNG, the median of the CV set is well converged to the original TNG300-1 simulation (which has comparable initial mass resolution as CAMELS).
        For Simba, the median of the CV set is not converged to the original Simba simulations for $N_\textnormal{HI} < 10^{14.5}$ cm$^{-2}$ and diverges more for lower $N_\textnormal{HI}$. 
        However, the divergence between CAMELS and original Simba is on par with the maximum variation from the CV set.
        This variation comes from the fact that the lowest \Lya~forest CD statistics in Simba are extremely sensitive to the AGN jet feedback and the CAMELS-Simba simulations are not volume-converged due to the long-range impact of jets \citep[seen in][and Gebhardt et al. submitted]{Borrow:2020}. 

        As shown in Figure \ref{fig:CDD}, the lowest column densities are particularly susceptible to variations in the AGN jet feedback mode. 
        Since the AGN jets can reach the outskirts of galaxies before recoupling to the heating and cooling of the simulation (distances on the order of $\sim 10$ kpc) they are able to carry most of their energy beyond galaxy scales.
        This energy, in many cases, is then enough to reach beyond the CGM and halo scales to reach the IGM.
        That heating will be more effective in lower density gas which could cause the divergence we see approaching lower $N_\textnormal{HI}$.
        In fact, the fiducial CAMELS Simba simulation has an overall hotter IGM than that of original Simba. 
        For the lowest values of $n_\textnormal{HI}$, corresponding to more diffuse absorbers, this temperature difference is hotter by a factor of $\sim 1.7$. 
        Since in the diffuse IGM $N_\textnormal{HI} \propto T^{-0.7}$ this temperature difference leads to $N_\textnormal{HI}$ values in original Simba that are $\sim 1.5$ times larger than that in CAMELS Simba. 
        Thus the temperature difference is enough to explain the offset between CAMELS and original Simba at the lowest column densities. 
        
        The origin of the temperature difference is almost certainly from the number of BHs producing jet feedback in the simulation box.
        This number will be affected by the random seed due to different halo mass functions being produced in each CAMELS box. 
        A few more massive halos in the box due to a different random seed will result in an increased number of SMBHs large enough to produce impactful long-range jet feedback. 
        A greater number of massive halos with strong AGN jets can be compounded by the small CAMELS box. 
        The periodic boundary conditions in combination with extremely far impacting jets could also result in the energy of these jets being trapped within the box from the point of view of the host halo rather than being dispersed at some distance as it would in reality.
       
        This idea is further motivated by the variance of the TNG CDD due to initial conditions. 
        The maximum variance of the CDD due to different random seeds seen in the CV set is significantly lower than the variance seen in Simba. 
        We saw repeatedly that modifying the AGN feedback parameters explored in the CAMELS TNG suite does not affect the \Lya~forest statistics implying that the impact of TNG's AGN feedback might have minimal impact on the forest (apart from the radiative mode which mimics UVB effects). 
        Different AGN feedback parameters may affect the forest or the same AGN parameters explored herein may affect the forest in a larger simulation box.
        
        The minimal impact seen from the AGN feedback implies that the variance of the TNG CDD due to the random seed is likely to be a better indicator of the effect of the random seed on just the population of \Lya~absorbers (i.e.\ whether or not the box size used herein is conducive to a representative population of \Lya~absorbers).
        Since the variance of TNG at low $N_\textnormal{HI}$ is so much lower than that of Simba, it is likely that another factor (e.g.\ AGN feedback) is affecting the Simba variance.
        The large variance seen at higher $N_\textnormal{HI}$ is similar in both simulations and is due to the rarity of those absorbers.
        Different random seeds will produce more high CD absorbers than others and this will be especially variable in the small box sizes of CAMELS. See Appendix \ref{Appendix:CVandConverge} for additional discussion on cosmic variance and convergence.

        \subsection{Broader Picture}

        In \citet{Burkhart_2022} the differences between the Illustris and TNG \Lya~forest statistics due to different AGN feedback models was found to be difficult to disentangle from effects of the UVB.
        The CDD in Illustris was able to match that of TNG by a re-scaling of the assumed UVB and the $b$-value distribution showed no difference between the two simulations.
        Those results implied that current observations of the CDD and $b$-values are unable to constrain the different AGN feedback models in Illustris and TNG.
        However, \citet{Burkhart_2022} found that the \Lya~flux statistics such as the flux PDF and the 1D flux power spectrum might show observable differences between the two simulations.

        Another recent study, \citet{Khaire:2023}, comparing Illustris and TNG found similar results with the CDD and $b$-value distribution showing no observable differences but the \Lya~flux power spectrum potentially showing observable differences at the highest $k$ values (on small physical scales).
        Further exploration of the flux power spectrum (particularly around $z=2$ to 3) could reveal a dependence on AGN feedback that may become important for certain studies. 
        For example, in the PRIYA simulation suite the effects of AGN feedback were very small ($<0.1\%$) in the 120 Mpc box and slightly larger in 25 Mpc box but only due to cosmic variance \citep{Bird:2023}.
        However, the AGN feedback model in those simulations resembles closely that of TNG and it is unclear if the Simba model would have more of an effect.

        Despite the Illustris and TNG simulation results implying the observed \Lya~forest CDD cannot be used to constrain AGN feedback, further work done in \citet{Tillman:2023} found that AGN feedback in Simba might be observable. 
        The changes in the Simba simulation CDD when removing AGN jet feedback showed potentially observable differences in the intermediate CD range ($N_{HI} \sim 10^{13}$ cm$^{-2}$) even when allowing the UVB strength to vary to obtain the best fit.
        Conducting a similar analysis to the one seen in \citet{Tillman:2023} on the simulations in this work, we analyze what parameter variations lead to observable changes that can be disentangled from degenerate effects of the UVB.
        We conduct fits in the CD range of $10^{13} < N_{HI} < 10^{14.5}$ cm$^{-2}$ as this is the range that observation and simulation data is most robust. Reasonable variations of the chosen fitting range does not change the main results of this work. We utilize the observational error bars as weights in our fits and we assume Gaussian distributed random variables.

        Figure \ref{fig:CDD_UVBcorr} is the same as Figure \ref{fig:CDD} but now allowing the strength of the UVB to vary in order to find the best fit to the observed data. 
        Each fit has two free parameters, the value of the feedback parameter being varied (\ASN{} or \AAGN{}) and the factor the assumed UVB model is multiplied by (\citet{Haardt:2012} for Simba and \citet{Faucher-Giguere:2009} for TNG).
        The factor by which the UVB is multiplied and the reduced $\chi^2$ value for a given parameter variation can be found in Table \ref{tab:UVBcorr} in Appendix \ref{Appendix:BestFits}.
        The most noticeable differences (relative to the observational error bars) in the CDD, when accounting for UVB degeneracy, can be seen for variations in the Simba AGN jet speed \AAGN{2} and the Simba SN wind speed \ASN{2}.
        A noticeable flattening of the CDD can be seen for increases in \AAGN{2} and decreases in \ASN{2}. 
        
        In particular, the most dramatic differences can be seen when increasing \ASN{2} from the fiducial value.
        We have already confirmed that the effects on the forest due to \ASN{2} are due to the link between stellar feedback and the suppression of SMBH seeding and growth.
        The dramatic reduction in the number of AGN producing jet feedback results in a much steeper CDD that resembles that of TNG.
        This effect resembles that of turning off the AGN jets in Simba seen in \citet{Tillman:2023}, but manifests more prominently due to the fact that the AGN feedback has a larger impact on the forest in the CAMELS-Simba box than in the original Simba run (an idea we discussed to explain the CV set results).
        The results from Figure \ref{fig:CDD_UVBcorr} imply that the effects from the AGN feedback in Simba could be unique enough to disentangle from the assumed UVB model and that observations of the \Lya~CDD could be used to constrain said feedback.
        In future work, we plan to explore the \Lya~forest flux statistics in the context of Simba's AGN feedback to determine if, similar to TNG vs.\ Illustris, the flux power spectrum might be used to constrain AGN jets.

        Finally, we will briefly discuss the discrepancy seen between the observed $b$-value distribution and the distribution predicted from simulations. 
        Simulations have consistently under-predicted the number of high $b$ absorbers (at low-$z$) with an observed mean lying around 30 km/s and the simulations predicting a mean around 20 km/s.
        While we cannot directly compare the CAMELS predicted $b$-value distribution to observations due to numerical resolution limitations, we can acknowledge the role that faster AGN jets and slower SN wind speeds play roles in broadening the $b$-value distribution. 
        The heating causing the changes in Figure \ref{fig:bvalue} originates from AGN jet shocks and dispersion of the heat the jets carried into the IGM when said jets re-couple hydrodynamically to the gas in the box.
        While this heating alone cannot explain the discrepancy between the observed and simulated $b$-value distribution, it could act as a partial solution.
        Then the amount of missing turbulence required to fully resolve the statistic would be slightly less than what has been predicted by previous studies \citep[e.g.\ by][]{Viel:2017, Gaikwad:2017viper, Bolton:2021}.

\section{Conclusion}\label{s:Conclusion}

    In this study we use the CAMELS 1P set of simulations to explore effects on \Lya~forest statistics due to parameter variation in the Simba and IllustrisTNG feedback models.
    We find that, in Simba, all four feedback parameters that were varied - the AGN momentum flux, AGN jet speed, SN mass loading factor, and SN wind speed - have clear effects on the \Lya~forest CDD and $b$-value distribution when ignoring degeneracy due to the assumed UVB. 
    When accounting for the plausible effects of the UVB on the CDD, we found that the AGN jet speed and the SN wind speed showed noticeable differences in the CDD when varied. 
    
    For TNG, none of the AGN feedback parameter variations explored in the CAMELS simulation suite affected the \Lya~forest statistics in a discernible way. 
    The only effect AGN feedback in TNG appeared to have on the \Lya~forest in this study was due to the radiative mode which strengthens the UVB.
    Varying stellar feedback parameters in TNG (SN energy per unit SFR and SN wind speed) showed minimal effects on both the CDD and the $b$-value distribution. 

    In both Simba and TNG we found that adjusting stellar feedback parameters had an indirect effect on the \Lya~forest. 
    Stellar feedback in these simulations has the power to limit BH seeding, growth, and feedback to the point of having an impact on the forest statistics. 
    The Simba simulation stellar feedback parameters had a significantly more dramatic effect than that seen in TNG, which follows from the fact that AGN feedback in Simba has a dramatic affect on the IGM whereas TNG AGN does not.
    Additionally, the Simba SMBH seeding prescription depends on stellar mass rather than halo mass which results in the stellar feedback in Simba having a large effect on the number of SMBHs in the box.
    We explored the extent to which stellar and AGN feedback affected the BHs in the simulation by looking at Eddington ratios and SMBH accretion rates for relevant AGN feedback modes, the number of SMBHs in each mode, and the number of SMBHs in the box.
    The main conclusions of our work are as follows:

    \begin{itemize}
        \item As found in previous work, Simba's AGN jet feedback plays a dominant role in the low-redshift \Lya~forest as the jets heat gas well outside of halos and into the IGM. 
        The distance the heat is transported is more important than the amount of gas ejected.
        The AGN jet feedback can change the shape of the CDD by flattening it.
        \item The $b$-value distribution in Simba is broadened by heating from stronger AGN feedback (faster jets and to a lesser degree higher momentum flux). 
        Heating from AGN \textit{jet} feedback (clearly seen in Figure~\ref{fig:Simba_proj_AAGN}) may be a partial solution in resolving the discrepancy between the observed and simulated $b$-value distribution.
        \item In agreement with previous work, we find the TNG AGN feedback model has minimal effect on the IGM and thus the \Lya~forest. 
        The AGN radiative mode affects the IGM which results in small changes when varying stellar feedback, but the effects of the radiative mode are largely degenerate with that of the UVB.
        \item  Stellar feedback plays a role in SMBH growth suppression and thus should be considered as a degenerate parameter along with the strength of the  UVB and AGN feedback when analyzing the low-$z$ \Lya~forest statistics.
    \end{itemize}

    The CAMELS-Simba simulations for parameter variations closest to the fiducial values produced the best fits to the observed CDD overall when allowing the strength of the UVB to vary.
    The CAMELS-TNG simulations could not produce a CDD that matches observations for any parameter variation. 
    Both CAMELS-TNG and CAMELS-Simba produced $b$-value distributions with mean values around 20 km/s which is too low relative to observations.
    Stronger AGN jet feedback in CAMELS-Simba (via faster jet speeds or lower SN wind speeds) broadened the $b$-value distribution implying heating from AGN jets could be a partial solution to resolving the predicted $b$-values from simulations to the observed $b$-value distribution.
    However, non-thermal broadening is likely necessary to fully resolve the $b$-values distribution via un-resolved or un-modeled forms of turbulence in simulations or via other instability mechanisms such as pressure from cosmic rays.

    Understanding the interplay between degenerate factors that affect the neutral hydrogen in the IGM is the first step in constraining these mechanisms with observational data.
    The next step to determine the extent of the degeneracy between the UVB, stellar feedback, and AGN feedback in these sub-grid models is to employ machine learning techniques to the CAMELS project simulations to determine best fits.
    Many other studies have noted degeneracies between UVB and AGN feedback effects on the forest, and it is clear that constraining feedback models to additional observables will be necessary to unravel the relationship \citep{Burkhart_2022,Tillman:2023,Mallik:2023,Khaire:2023}.
    This will not only help simulations reporoduce a wider range of observables, but will aid in the construction of new more physical feedback models.

    Finally it is important to acknowledge that additional factors not discussed in this study likely affect the large-scale environment of the IGM. 
    For example, the role of cosmic rays in determining the amount and distribution of HI in the IGM is relatively unconstrained \citep{Lacki:2015,Leite:2017,Butsky:2023}. 
    Exploring other cosmological and astrophysical mechanisms that affect the IGM in addition to the UVB and galactic feedback will be essential in fully resolving the discrepancies between the observed and simulated \Lya~forest.
    In future work we plan to do a more thorough analysis of individual absorbers between simulations to determine which factors apart from AGN feedback are defining the forest statistics.
    We also plan to analyze the \Lya~flux power spectrum in the Simba simulation for variations in the AGN feedback models to determine the extent to which those effects might be observable or affect constraints are dark matter properties.

\begin{acknowledgments}
    All authors acknowledge support for this work by  NASA ATP 80NSSC22K0823.	
    MTT thanks the Simons Foundation and the Center for Computation Astrophysics at the Flatiron Institute for providing the computational resources used for this analysis. 
    MTT thanks Mahdi Qezlou for helpful conversations regarding proper usage and trouble shooting of the \textit{fake-spectra} package. 
    MTT thanks Doug Rennehan for insightful conversations about the difference between the Simba and CAMELS CDD due to cosmic variance.
    BB is grateful for additional support by NSF grant 2009679, the Packard Fellowship, and the Sloan Fellowship. 
    Part of this work was completed during the KITP workshop Galevo 23.
    DAA acknowledges support by NSF grants AST-2009687 and AST-2108944, CXO grant TM2-23006X, Simons Foundation Award CCA-1018464, and Cottrell Scholar Award CS-CSA-2023-028 by the Research Corporation for Science Advancement.
    GLB acknowledges support from the NSF (AST-2108470, XSEDE grant MCA06N030), NASA TCAN award 80NSSC21K1053, and the Simons Foundation (grant 822237) and the Simons Collaboration on Learning the Universe. 
    SH acknowledges support for Program number HST-HF2-51507 provided by NASA through a grant from the Space Telescope Science Institute, which is operated by the Association of Universities for Research in Astronomy, incorporated, under NASA contract NAS5-26555. 
    SB acknowledges additional support from NSF AST-2107821.
\end{acknowledgments}

\appendix

    \section{Cosmic Variance and Convergence}\label{Appendix:CVandConverge}

\begin{figure*}
            \centering
            \includegraphics[width = 0.7\linewidth, trim=0.0cm 0.0cm 0.0cm 0.0cm, clip=true]{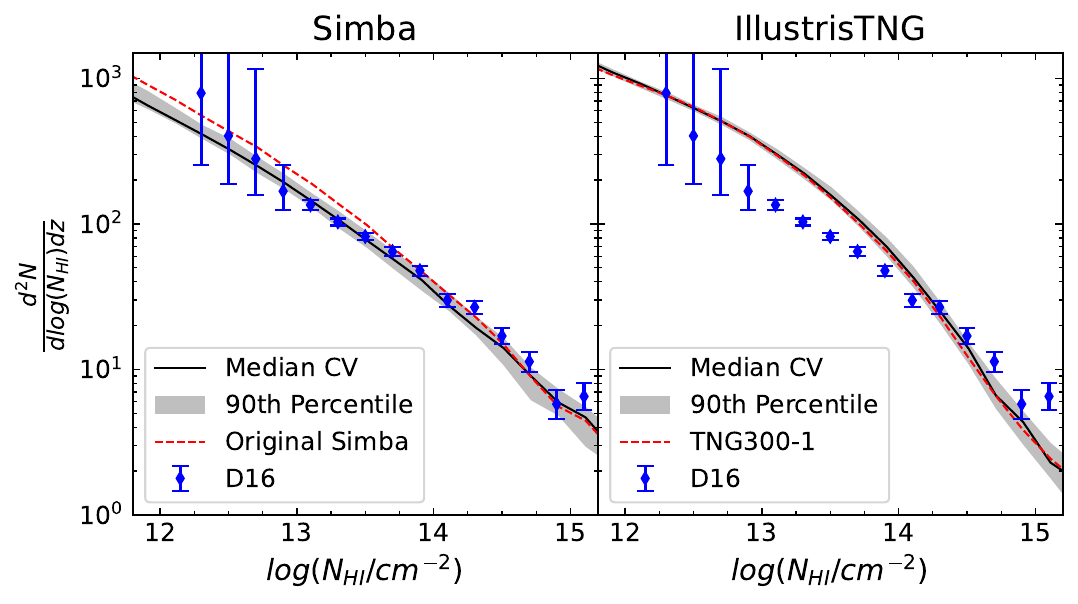}
            \caption{The CDD for all the CAMELS CV set simulations for both the SIMBA and IllustrisTNG suites. The solid black lines are the median of the CAMELS CV CDDs, the shaded region is the 90th percentile range of the CV CDDs, and the dashed red lines are the original Simba and TNG300-1 simulations. Variations in the CDDF due to varying initial conditions results in a \textit{maximum} shift in the overall normalization by ~0.25 dex. Both the original Simba and TNG300-1 simulations are largely contained within the allowed variation of the CDD due to cosmic variance.}
            \label{fig:CDD_CV}
        \end{figure*}

        We examine the effect that different initial random seeds have on the CDD and $b$-value distribution. 
        By understanding to what extent the results may change based on initial conditions we can better examine what is an effect of feedback vs.\ a sampling effect.
        Figure \ref{fig:CDD_CV} shows the range of variation in the CDD from the CV set. 
        Displayed is the median CDD from all the CV simulations for each suite. The shaded region represents the range of variation in the CDD for the CV set.
        Also displayed are the original Simba and IllustrisTNG CDDs. 
        We compare to the TNG300-1 simulation as it has a comparable mass resolution to that of CAMELS. 
        The CAMELS TNG results are converged to the original TNG results however the original Simba results diverge from the CAMELS Simba results at lower CDs.
        The Simba divergence is slightly larger than the allowed CV range. Previous work had already found that the TNG CDD is converged for the resolution explored herein \citep{Burkhart_2022}. However, for Simba it is unclear if even the original simulation produces a converged CDD due to the lack of higher resolution simulations for comparison. Since the AGN feedback model in the Simba simulations has such a dramatic effect on the predicted CDD, it is possible that a higher resolution simulation may be necessary to properly constrain said feedback model.

\begin{figure*}
            \centering
            \includegraphics[width = 0.8\linewidth, trim=0.0cm 0.0cm 0.0cm 0.0cm, clip=true]{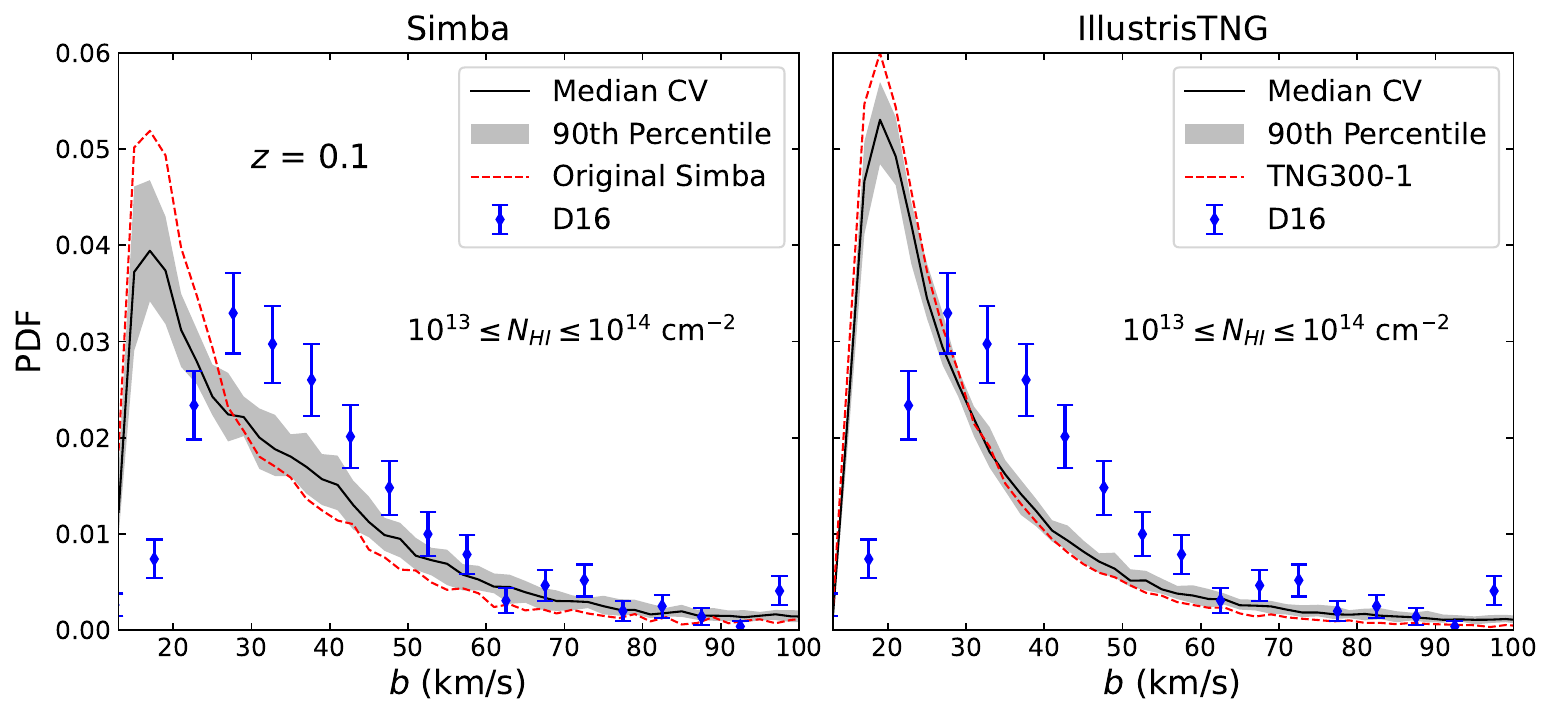}
            \caption{Same as Figure \ref{fig:CDD_CV} but instead for the $b$-value distribution. The blue points are observational data from \citet{Danforth:2016}. The $b$-value distributions are calculated only for absorbers in the column density range corresponding to the observational data as written in the figure panels. The x-axis range ends at 100 km/s to avoid blends of multiple lines and continuum fitting errors. }
            \label{fig:bvalue_CV}
        \end{figure*}

        Figure \ref{fig:bvalue_CV} shows the median $b$-value PDF from the CV set as well as the range of variation. 
        The variation in the PDF is higher at lower $b$-values. 
        This is likely a result of the fitting algorithm as low $b$-value absorbers are harder to fit via an automated method and are more likely to be noise from a previous fit rather than a unique absorber.
        The predicted $b$-value distributions from CAMELS are not converged for any CDs range, but diverge more particularly for lower CDs ($N_{HI} < 10^{14}$ cm$^{-2}$). This was determined prior to this work through a resolution study with the IllustrisTNG simulations \citep{Burkhart_2022}. As mentioned throughout, we do not directly compare the CAMELS $b$-value distributions to observations due to the lack of convergence. We expect the CAMELS-TNG results to be close to convergence, but both CAMELS suites lack the mass resolution necessary for confident comparisons. However, we include the \citet{Danforth:2016} observational data in Figure \ref{fig:bvalue_CV} for reference. 
        
        The range of column densities shown in Figure \ref{fig:bvalue_CV} corresponds to that of the \citet{Danforth:2016} observational data ($10^{13} < N_{\textnormal{HI}} < 10^{14}$ cm$^{-2}$) and the CAMELS b-value distribution in this range is closer to convergence than if we were to include the $N_\textnormal{HI} < 10^{13}$ cm$^{-2}$ data. Furthermore, if we only look at absorbers with $N_\textnormal{HI} > 10^{14}$ cm$^{-2}$ then the CAMELS predicted $b$-value distributions do converge to the original Simba and TNG simulations exemplifying the importance of mass resolution for low CD absorber statistics. Regardless of convergence, none of the simulations produce a distribution that looks like the observed data, in fact it appears the lower resolution (not converged) distributions actually produce a closer match than the higher resolution results. This is consistent with previous studies' findings that simulations consistently under predict b-values when compared to observations indicating a lack of heating or turbulence in the simulated low-$z$ \Lya~forest \citep{Viel:2017, Gaikwad:2017viper, Bolton:2021, Burkhart_2022}.

    \section{UVB Corrections and Best Fits}\label{Appendix:BestFits}

        We conduct a least squares fit to find the UVB correction factor required for the best fit of the CDDs predicted by the CAMELS 1P simulations explored herein to the D16 observational data.
        The fitting procedure is heavily based on that used in \citet{Tillman:2023}.
        For our fitting procedure, we assume Gaussian distributed random variables.
        We conduct this fit within a CD range of $N_{\rm HI} = 10^{13}$ cm$^{-2}$ to $10^{14.5}$ cm$^{-2}$ as this is where both the simulation data and observations are most robust. Reasonable variations of this fitting range does not change the main results of this work.
        Excluding lower CD values from the fit due to observational scarcity will not affect the main results of these fits due to the high observational error bars below $N_{HI} \approx 10^{13}$ cm$^{-2}$.
    
        We closely follow a UVB correction method as outlined in \citet{Kollmeier:2014} which uses the approximation that $N_{\rm HI} \propto 1/\Gamma_{\rm HI}$ where $\Gamma_{\rm HI}$ is the hydrogen photoionization rate.
        The relation works since the low redshift \Lya~forest can be well approximated as an optically thin region in photoionization equilibrium.
        This method breaks down when absorbers are no longer optically thin but is well converged for CDs explored herein and can be applied in post-processing.
    
        For the reduced $\chi^2$ ($\chi^2_R$), the number of degrees of freedom is the number of observational points being fit with two variable parameters: the UVB correction factor and the feedback parameter varied in CAMELS. 
        For several parameter variations of CAMELS-Simba, the value for $\chi^2_R$ is below 1 implying the CDD is over fitted (this could be partially fixed by removing the large observational error bar points at low CDs). However, many variations including the fiducial results produce $\chi^2_R$ values close to 1 exemplifying the remarkable fit to the observed data predicted by CAMELS-Simba.
        The CAMELS-Simba fits are better than the original Simba fits found in \citet{Tillman:2023} due to a further flattening of the CDD in CAMELS-Simba that manifests due to box-size. 
        This effect was discussed in Section \ref{s:Discussion}.
        
        Recent studies have found hydrogen photoionizing values at $z=0.1$ that are $\sim$ 1.77, 1.78, 2.56, and 1.74 \citep[for][respectively]{Gaikwad:2017, KhaireUVB:2019, Puchwein:2019, FG:2020} times stronger than the \citet{Haardt:2012} values. 
        These values can go as high as $\sim 5$ times stronger when allowing the escape fraction of HI ionizing photons from galaxies to vary \citep{Khaire+Srianand:2015}.
        While UVB correction factors lower than \citet{Haardt:2012} found for many of the CAMELS-Simba best fits are disfavored by these more recent UVB model studies, the UVB at low-$z$ is still not well constrained which is why it is often corrected out of \Lya~statistics in studies like this one.

        \begin{table}[]
            \centering
            \begin{tabular}[t]{|l|l|l|}
            \hline
            \multicolumn{3}{c}{\textbf{Simba}} \\
            \hline \hline
            Parameter & $\times \Gamma _{HI}^{a}$ & $\chi^2_R$ \\
            \hline
            \AAGN{} = 1 & 0.94 & 1.70 \\
            \hline
            \AAGN{1} = 0.25 & 1.57 & 6.15 \\
            \AAGN{1} = 0.33 & 1.51 & 3.40 \\
            \AAGN{1} = 0.44 & 1.24 & 4.26 \\
            \AAGN{1} = 0.57 & 1.24 & 1.45 \\
            \AAGN{1} = 0.76 & 1.15 & 1.40 \\ 
            \AAGN{1} = 1.32 & 0.89 & 0.24 \\
            \AAGN{1} = 1.74 & 0.95 & 1.06 \\
            \AAGN{1} = 2.30 & 0.82 & 1.73 \\ 
            \AAGN{1} = 3.03 & 0.86 & 1.61 \\
            \AAGN{1} = 4.0 & 0.96 & 0.65 \\
            \hline
            \AAGN{2} = 0.50 & 2.05 & 7.30 \\
            \AAGN{2} = 0.57 & 1.71 & 4.52 \\
            \AAGN{2} = 0.66 & 1.6 & 4.50 \\ 
            \AAGN{2} = 0.76 & 1.41 & 2.69 \\
            \AAGN{2} = 0.87 & 1.11 & 1.12 \\
            \AAGN{2} = 1.15 & 0.88 & 2.02 \\
            \AAGN{2} = 1.32 & 0.63 & 0.66 \\
            \AAGN{2} = 1.52 & 0.56 & 1.0 \\
            \AAGN{2} = 1.74 & 0.49 & 1.33 \\
            \AAGN{2} = 2.0 & 0.28 & 2.85 \\
            \hline
            \end{tabular}
            \hfill
            \begin{tabular}[t]{|l|l|l|}
            \hline
            \multicolumn{3}{c}{\textbf{Simba}} \\
            \hline \hline
            Parameter & $\times \Gamma _{HI}^{a}$ & $\chi^2_R$ \\
            \hline
            \ASN{} = 1 & 0.94 & 1.69 \\
            \hline
            \ASN{1} = 0.25 & 1.60 & 4.32 \\
            \ASN{1} = 0.33 & 1.42 & 2.92 \\
            \ASN{1} = 0.44 & 1.43 & 4.45 \\
            \ASN{1} = 0.57 & 1.21 & 1.28 \\
            \ASN{1} = 0.76 & 1.08 & 2.29 \\
            \ASN{1} = 1.32 & 0.94 & 0.74 \\
            \ASN{1} = 1.74 & 0.88 & 1.36 \\
            \ASN{1} = 2.30 & 0.96 & 0.99 \\
            \ASN{1} = 3.03 & 1.05 & 1.94 \\
            \ASN{1} = 4.0 & 1.18 & 2.74 \\
            \hline
            \ASN{2} = 0.50 & 0.35 & 0.84 \\
            \ASN{2} = 0.57 & 0.40 & 1.14 \\
            \ASN{2} = 0.66 & 0.58 & 1.08 \\
            \ASN{2} = 0.76 & 0.65 & 1.30 \\
            \ASN{2} = 0.87 & 0.76 & 0.86 \\
            \ASN{2} = 1.15 & 1.26 & 3.97 \\
            \ASN{2} = 1.32 & 1.76 & 5.14 \\
            \ASN{2} = 1.52 & 2.33 & 9.44 \\
            \ASN{2} = 1.74 & 2.79 & 17.2 \\
            \ASN{2} = 2.0 & 3.24 & 20.1 \\
            \hline
            \end{tabular}
            \hfill
            \begin{tabular}[t]{|l|l|l|}
            \hline
            \multicolumn{3}{c}{\textbf{IllustrisTNG}} \\
            \hline \hline
            Parameter & $\times \Gamma _{HI}^{b}$ & $\chi^2_R$ \\
            \hline
            \ASN{} = 1 & 2.12 & 40.9 \\
            \hline
            \ASN{1} = 0.25 & 1.76 & 39.0 \\
            \ASN{1} = 0.33 & 1.71 & 36.2 \\
            \ASN{1} = 0.44 & 1.85 & 38.1 \\
            \ASN{1} = 0.57 & 1.99 & 42.1 \\
            \ASN{1} = 0.76 & 2.06 & 43.2 \\
            \ASN{1} = 1.32 & 2.13 & 45.0 \\
            \ASN{1} = 1.74 & 2.40 & 44.1 \\
            \ASN{1} = 2.30 & 2.29 & 43.9 \\
            \ASN{1} = 3.03 & 2.32 & 47.1 \\
            \ASN{1} = 4.0 & 2.30 & 46.1 \\
            \hline
            \ASN{2} = 0.50 & 1.91 & 35.8 \\
            \ASN{2} = 0.57 & 1.83 & 37.1 \\
            \ASN{2} = 0.66 & 2.01 & 36.0 \\
            \ASN{2} = 0.76 & 1.96 & 40.1 \\
            \ASN{2} = 0.87 & 2.04 & 39.6 \\
            \ASN{2} = 1.15 & 2.1 & 48.4 \\
            \ASN{2} = 1.32 & 2.1 & 43.6 \\
            \ASN{2} = 1.52 & 2.17 & 49.8 \\
            \ASN{2} = 1.74 & 2.02 & 48.4 \\
            \ASN{2} = 2.0 & 2.09 & 48.9 \\
            \hline
            \end{tabular}
            \caption{UVB Corrections and Reduced $\chi^2$ for Figure \ref{fig:CDD_UVBcorr}.
            \\
            $^{\textnormal{a}}$ Factor by which the \citet{Haardt:2012} UVB is multiplied by at $z=0.1$.
            \\
            $^{\textnormal{b}}$ Factor by which the \citet{Faucher-Giguere:2009} UVB is multiplied by at $z=0.1$.}
            \label{tab:UVBcorr}
        \end{table}

\bibliography{mybib}{}
\bibliographystyle{aasjournal}
 
\end{document}